\documentclass{aa}  
\usepackage[varg]{txfonts}
\usepackage{natbib}
\usepackage{graphicx}
\usepackage{epstopdf}
\usepackage{color}
\usepackage{soul}
\usepackage{comment}
\usepackage{txfonts}
\usepackage[normalem]{ulem}

\begin{document}


\title{A constellation of SmallSats with synthetic tracking cameras 
to search for 90\% of potentially hazardous near-Earth objects
}

\titlerunning{A constellation of SmallSats to search for NEOs}

\author{Michael Shao, Slava G. Turyshev, Sara Spangelo, Thomas Werne, and Chengxing Zhai}
\authorrunning{M. Shao, S.G. Turyshev, S. Spangelo, T. Werne, C. Zhai}

\institute{Jet Propulsion Laboratory, California Institute of Technology\\ 
4800 Oak Grove Drive, Pasadena, CA 91109-8099, USA
}


%\begin{abstract}
\abstract{
We present a new space mission concept that is capable of finding, detecting, and tracking 90\% of near-Earth objects (NEO) with H magnitude of $\rm H\leq22$ (i.e., $\sim$140 m in size) that are potentially hazardous to the Earth. The new mission concept relies on two emerging technologies: the technique of synthetic tracking and the new generation of small and capable interplanetary spacecraft.  Synthetic tracking is a technique that de-streaks asteroid images by taking multiple fast exposures. With synthetic tracking, an 800 sec observation with a 10~cm telescope in space can detect a moving object with apparent magnitude of 20.5 without losing sensitivity from streaking. We refer to NEOs with a minimum orbit intersection distance of $< 0.002$ au as Earth-grazers (EGs),  representing typical albedo distributions. We show that a constellation of six SmallSats (comparable in size to 9U CubeSats) equipped with 10~cm synthetic tracking cameras and evenly-distributed  in  1.0~au heliocentric orbit could detect 90\% of EGs with $\rm H \leq 22~mag$ in $\sim$3.8 years of observing time. A more advanced constellation of nine 20~cm telescopes could detect 90\% of $\rm H=24.2~mag$ (i.e., $\rm \sim 50~m$ in size) EGs in less than $5$~years.}
%\end{abstract}


\date{Received 29 September 2016 / Accepted }



\maketitle
 
\section{Introduction}
\label{sec:intro}

The discovery and characterization of near-Earth objects (or NEO, which is an object that orbits the Sun and approaches or crosses the Earth's orbit) is motivated by several aspirations and concerns, including scientific research, planetary protection, and exploration efforts.  NEOs are believed to be remnants from the early evolution of the solar system and hence studies of their dynamics and chemical composition may offer important information about conditions  at that early epoch. The possibility that some NEOs could approach and even impact the Earth has motivated many observers worldwide to systematically search, catalog, and study the NEO population. Over the years, NASA has funded several NEO surveys to find such potentially hazardous objects (PHO).

 To date, approximately 12,300 NEOs have been discovered\footnote{The most recent information on the NEOs discovered may be obtained from the IAU's Minor Planet center at: {http://www.minorplanetcenter.net/}}, 820 of which have a diameter larger than 1~km. From that population, 1,492 objects (12.2\%) are classified as ``potentially hazardous'' with a minimum orbit intersection distance (MOID) of less than 0.05 astronomical units (au). A recent Report by the US National Research Council \citep{NRC2010} recognized the fact that objects smaller than 140 m in size are also capable of causing significant damage to the Earth. Estimates show \citep{Harris:2009,Harris:2011} that there are millions of asteroids with sizes ranging from 140~m down to 30~m 
that are still undetected, but those objects are large enough to cause major regional damage in the event of an Earth impact.  Early analysis of the object that entered the Earth's atmosphere over the Siberian wilderness near Podkamennaya Tunguska in 1908 estimated that its size was $\sim$70~m.
However, recent analyses \citep{Chyba-etal:1993,Boslough-Crawford:1997,Boslough-Crawford:2008} indicate that the object could have been substantially smaller, perhaps 30 to 50 m, causing much of the damage by exploding in the atmosphere and resulting in shock waves that devastated more than 2,000~km$^2$ of forest.  Accordingly, NEOs as small as 30--50~m in size could be highly destructive.  Among recent events, the Chelyabinsk meteor in 2013 had a diameter of only 17~m prior to entering the Earth's atmosphere. Therefore, in addition to the efforts of finding objects 140~m and larger, there is a need for detecting as many objects that are 30--50~m (and, perhaps, even smaller) as possible. The possibility of mining of near-Earth asteroids (NEAs, a subset of the NEOs) has also provided an impetus to private companies to perform a census of these objects to identify the most viable targets. 

Recognizing the threat that NEOs pose to life on Earth, the US Congress has passed the 2005 NASA Authorization Act\footnote{Specifically, the George E. Brown, Jr. Near-Earth Object Survey section of the NASA Authorization Act of 2005 (Public Law 109-155).}, where NASA was mandated by the end of year 2020 to detect, track, catalog, and characterize at least 90\% of NEOs brighter than $\rm H=22$~mag that could potentially impact the Earth. [To guide the search for faint objects one uses either the size of a NEO or its brightness. However, because of the large variation in the bimodal distribution of albedos, there is no simple relationship between H mag and size. In this paper, we have conducted simulations of a constellation of SmallSats to detect 90\% of H=22 mag NEOs and in one case H=24.2 mag NEOs, which the 2010 NRC report would refer to as 140~m NEOs and 50~m NEOs, correspondingly.]  The US Congress also asked the NRC in 2008 to form a committee to determine the optimum approach to doing so.  In 2010, after a detailed look at a number of alternative approaches, the NRC concluded that the goal of finding 90\% of H=22~mag NEOs by the year 2020 was an almost impossible achievement \citep{NRC2010}.  The NRC committee looked at various ground- and space-based options and found that the most viable approaches with the potential to complete the survey in the period of approximately 10 years were the space-based ones. Each of the corresponding missions would rely on a single spacecraft, would require an active decade-long observing campaign, and would cost over \$500M.  However, even these expensive missions will not guarantee completion of the survey by 2020. 

The 2010 NRC report emphasized that a combination of the space-based efforts together with an appropriate ground-based facility (such as LSST\footnote{The Large Synoptic Survey Telescope (LSST), for details, see {http://www.lsst.org/lsst/about}} or PanSTARRS\footnote{The Panoramic Survey Telescope and Rapid Response System (Pan-STARRS), {http://pan-starrs.ifa.hawaii.edu/public/}}) could be used to accelerate the survey. Such a combination could complete the survey well before 2030, perhaps as early as 2022. They reported that using a large ground-based telescope alone (for instance, LSST), even if this facility would be fully dedicated to NEO search, could not complete the Congress-mandated survey by the original 2020 deadline. In fact, it would take a decade-long dedicated campaign to complete this effort probably just before 2030.  Furthermore, the committee concluded that, despite associated launch risk and a more limited lifetime, a space-based option could be the fastest means to complete the survey.
 
Since the release of the 2010 NRC report we witnessed significant technology progress in several relevant areas that could result in a major paradigm shift in the search for NEOs. One such area is the technique of synthetic tracking of NEOs \citep{Shao-etal:2014}, which makes it technically feasible to perform NEO searches with small CubeSat-compatible cameras. Synthetic tracking of NEOs is an emerging technology that was recently developed and demonstrated by NASA's Jet Propulsion Laboratory (JPL) \citep{Shao-etal:2014,Zhai-etal:2014}.  When used from a ground-based facility, this new tracking technique provides an order of magnitude improvement in the ability to detect and track dim and fast-moving objects.  If implemented on a 10~cm CubeSat telescope in space with 800~sec observations, synthetic tracking of NEOs would result in a sensitivity to a 20.5~mag object at SNR=7, making it an ideal candidate technology to use on a CubeSat. 
 
Another technological advancement that enables the proposed small spacecraft mission architecture  is the rapid development, flight heritage, and technology maturation for small and capable spacecraft, namely CubeSats. A single unit (1U) CubeSat\footnote{For information on CubeSats, please visit {http://en.wikipedia.org/wiki/CubeSat} and  {http://cubesat.jpl.nasa.gov/}.}  is a type of miniaturized satellite that usually has a volume of exactly one liter (based on a $[10~{\rm cm}\times 10~{\rm cm}\times 10~{\rm cm}]$ form-factor), has a mass of no more than 1.33~kg and typically costs \$1--2M.  
In fact, we observe that a CubeSat-based mission architecture presents a viable alternative relative to the missions proposed in the NRC report. Modern CubeSats have benefited from decades of development efforts in miniaturization of many spacecraft technologies. Typically, modern CubeSats use commercial off-the-shelf (COTS) space-qualified hardware that can be purchased at a cost that is dramatically lower relative to that of conventional spacecraft components. The radically lower cost of a CubeSat-based architecture enables one to consider using multiple small spacecraft, each consisting of several CubeSat units, while conventional multi-spacecraft architectures would never be economically viable. [A good example of an interplanetary CubeSat-based mission is the Mars Cube One (MarCO) mission, which is now ready for launch, see {http://www.jpl.nasa.gov/cubesat/missions/marco.php}. MarCO consists of two identical 6U CubeSats with the total cost for both being $\sim$\$10.2M. Interestingly, the actual incremental cost of the 2nd 6U CubeSat was \$1.0M. Thus, a constellation of six MarCOs with a total mass of 40~kg would cost $\sim$\$15M.]

We present a new mission concept that uses a constellation of SmallSats (that are in form-factor comparable with the 9U CubeSats) equipped with a 10~cm synthetic tracking telescope to survey over 90\% of the population of H=22 mag Earth-grazing NEOs (with MOID of $\leq0.02$~au) in $\sim$3.8 years of observing time, all at a cost that is dramatically lower relative to any of the missions examined by the NRC. Furthermore, a second generation constellation of synthetic tracking SmallSats could survey 90\% of 50~m NEOs in less than 5 years. 

This paper is organized as follows: In Section~\ref{sec:sim90}, we present the details of our recent simulation developed to study a search for $\rm H=22$~mag NEOs using multiple SmallSats in heliocentric orbits whose form factor is comparable to 9U CubeSats. In Section~\ref{sec:case-studies}, we discuss several SmallSats-based mission scenarios that could accomplish the Congress-mandated NEO search in $\sim$3.8 years of observing time. We also discuss a second generation SmallSats constellation that could search for 90\% of NEOs with sizes down to 50~m in diameter.  In Section~\ref{sec:discussion}, we summarize and discuss the results. Appendix~\ref{sec:syn-track} summarizes the synthetic tracking technique, which leads to high detection sensitivity when implemented with a 10~cm optics on a CubeSat. Appendix~\ref{sec:cubesat} discusses the emerging capabilities of interplanetary CubeSats. 

\section{Simulation of a survey to find 90\% of NEOs}
\label{sec:sim90}

The 2010 NRC report examined combinations of a spacecraft and a dedicated ground-based NEO search facility  \citep{NRC2010}, although the idea of putting multiple spacecraft into a solar orbit was not examined, likely because of the prohibitively large cost of launching multiple $\sim$0.5-1.0 billion dollar spacecraft.  Furthermore, low-cost small spacecraft were not studied, likely because they were perceived as incapable of accomplishing this mission. 

Compared to the approaches in the NRC report, our strategy towards accomplishing the NEO search is different: In particular, as opposed to using a single spacecraft, we explored the advantages offered by constellations of multiple spacecraft relying on small telescopes and a synthetic tracking technique (see Appendix~\ref{sec:syn-track}) for this purpose. To examine the advantages of using multiple spacecraft, we developed a simulator that takes into account the current model for distribution of the NEOs, expected performance of the synthetic tracking camera on a CubeSat, and the 9U NEO CubeSat operational characteristics, as well as contributions from the anticipated zodiacal dust background, and so on. 

We simulated a number of NEO observing scenarios using multiple spacecraft in solar orbits. The methodology of our simulation was designed to mimic the survey simulations discussed in the NRC 2010 report, with the exception of using mission architectures consisting of multiple spacecraft.  

\begin{figure}[ht]
%\epsscale{0.45}
%\plotone{fig7a.eps}
%\plotone{fig7b.eps}
\centering\includegraphics[width=7.5cm]{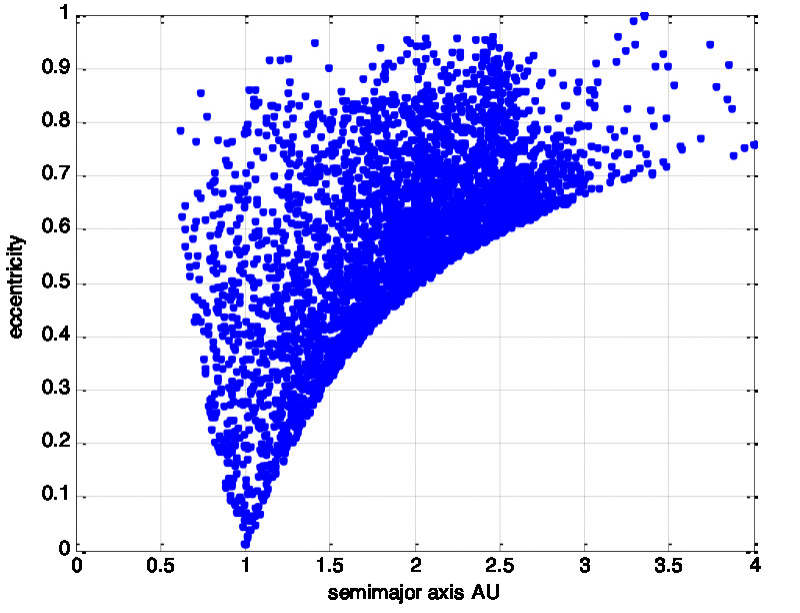}
\centering\includegraphics[width=7.5cm]{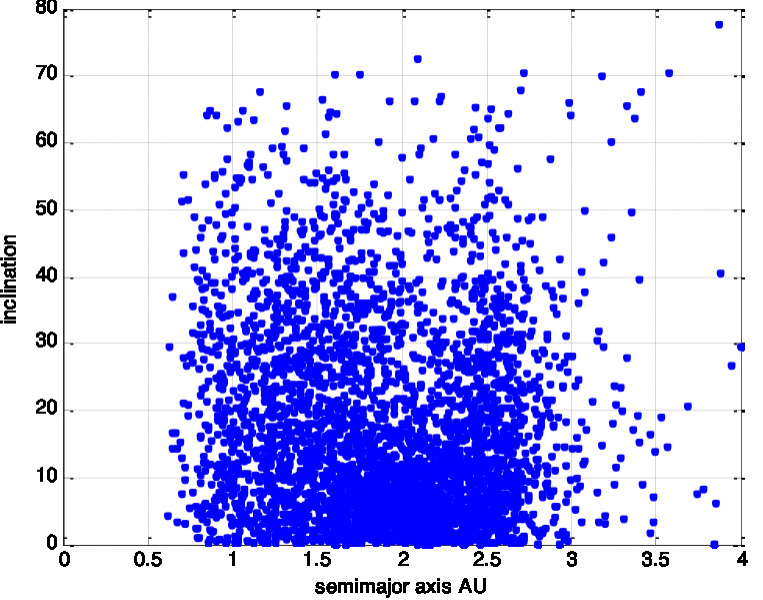}
\caption{Orbital distribution of $\sim$10,000 ``impactors'' found within the parent distribution taken from  \citep{Granvik-etal:2016}.
\label{fig:7}}
\end{figure}

We used the published debiased orbital and absolute magnitude distribution of NEOs  \citep{Granvik-etal:2016} that accounts for correlations in various orbital elements of NEOs. In the NEO population we took the NEO orbital parameters from \citep{Granvik-etal:2016}  for semi-major axis, eccentricity and inclination, but randomized the time of periastron. The Granvik population calculates the MOID, and for the Earth grazers, we chose only the NEOs with MOID~$< 0.01$~AU but chosen so that there were approximately $8,000-12,000$ objects. These objects are the ones most likely to impact the Earth at some time in the future, we call them Earth grazers (EG). The distribution of semi-major axis, eccentricity, and inclination for EGs are shown in Figure~\ref{fig:7}.

The SmallSat constellations were calculated using the following approach: Although each of the $N$ spacecraft could have an arbitrary orbit, we first examined those constellations where all the spacecraft had the same semi-major axis and inclination, but were uniformly distributed around the Sun.  The simulation then stepped  through $N$ to simulate constellations with variable numbers of spacecraft, typically considering scenarios up to 20 years. The time step for the simulation was $\sim$810~sec (representing $800$~sec integration time at one field of view (FOV) and 10 sec slew to next field). The time step was selected to represent the combination of the integration time and the time needed to slew the telescope to an adjacent FOV.  

At each time step, the telescope would point at a particular part of the sky.  Every EG was checked to determine if and when it was in the camera's FOV. If it was, the apparent magnitude of the EG was calculated, given its H magnitude, distance from the Sun and distance from the CubeSat and the phase angle (the angle between the telescope and the Sun as seen from the EG). The telescopes scanned the sky in an ``orange peel'' pattern and always satisfied the $65^\circ$ Sun-avoidance constraint. Currently, we do not simulate the loss of sensitivity when the EG image would be streaked, because it is not relevant with synthetic tracking. 

For a search of $\rm H=22$~mag EGs we look at a 10~cm aperture camera with a limit magnitude of 20.49~mag similar to many existing and near-future ground-based observatories (CSS\footnote{The Catalina Sky Survey (CSS), {http://www.lpl.arizona.edu/css/}} and ATLAS, PTF\footnote{The Palomar Transient Factory, {http://www.ptf.caltech.edu/iptf}}) not as sensitive as EG surveys with larger telescopes (i.e., PanSTARRS, LSST). This camera is also less sensitive than all of the space missions mentioned in the NRC report. With a 4K sensor and 3.3{\arcsec} pixels the FOV$\sim$14 $({}^\circ)^2$ is similar to the IR space telescopes in \citep{NRC2010}.

\section{Case studies}
\label{sec:case-studies}

We first validated our modeling approach and assumptions by reproducing the results for the IR telescope in Venus-trailing orbit from the 2010 NRC report (i.e., Section 5.1 in \citep{NRC2010}). We then investigated the trade-offs between telescope FOV, sky coverage rates, and survey time to verify the choice of the proposed aperture design for the CubeSat and scanning strategy.  We also explored different constellation sizes and heliocentric orbits to identify optimal constellation parameters that could determine the optimal trade-off between survey size. Finally, we investigate the ability of a constellation of SmallSats that relies on the CubeSat-derived technology to detect 90\% of smaller EGs with sizes down to 50~m in diameter. Below we present all these special cases and discuss results obtained.

\subsection{Case 1: Comparison to the proposed IR telescope on a Venus-like orbit}
 
To verify our simulation approach, we first aim to replicate the results for the NRC-proposed IR telescope in a Venus-like orbit. Our objective was to verify that our simulation was consistent with those conducted for the 2010 NRC study.  The information for this mission was obtained from publicly available resources\footnote{See information on the Sentinel mission: {http://sentinelmission.org/}}  and email correspondence [H. Reitsema, private communication (2014)]. The IR telescope on a Venus-like orbit is sufficiently sensitive to detect a 140~m EG from a distance of 0.6~au.  For a telescope at a distance of 0.7~au from the Sun, such a sensitivity yields a limiting magnitude of 21.5 mag or $\sim$1 mag more sensitivity than the telescope on our proposed CubeSat mission. The IR telescope would integrate for a total of 180 sec over six 30 sec exposures followed by a 60 sec slew/settle period. These six exposures will be used to remove cosmic ray events before co-adding the frames.  

The scanning strategy consists of revisiting areas of the sky $\sim$1 hr apart to confirm the detection of moving objects. We implemented an orange peel scanning pattern where the FOV of the CubeSat camera was properly taken into account. Assuming the FOV of 11~$({}^\circ)^2$ (similar to that of the Sentinel mission\footnote{http://sentinelmission.org/sentinel-mission/ sentinel-data-sheet/}), we modeled each observation as two observations of 240 sec including slew times separated by 1~hr.  When observing an EG that is at phase angle of 90${}^\circ$ from the Sun, only half of the surface is in sunlight. In the visible band, the apparent brightness of an EG at 90${}^\circ$ phase angle is approximately one third of its brightness at opposition. This phase angle effect was turned off in our simulation of the IR telescope.
 
\begin{figure}[t]
%\epsscale{0.5}
%\plotone{fig8.eps}
\centering\includegraphics[width=7.0cm]{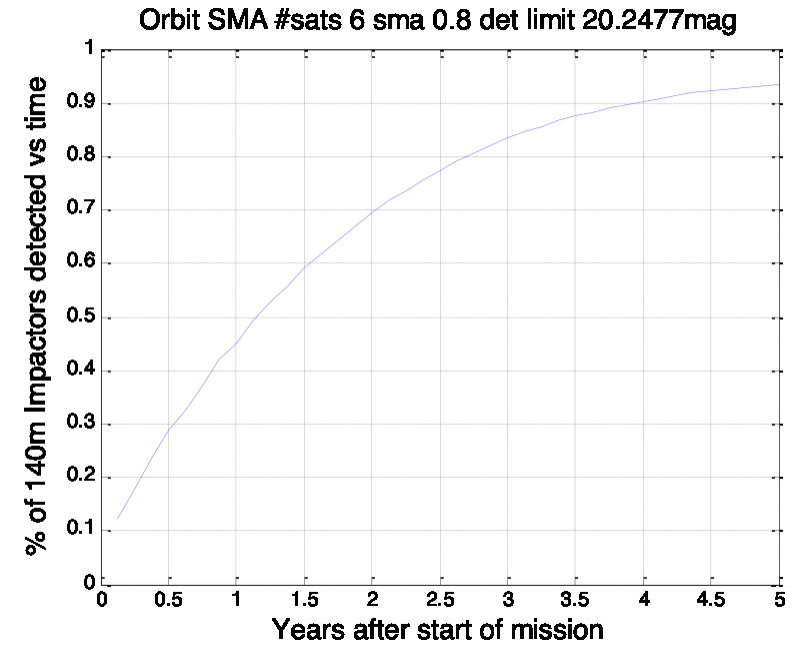}
\caption{Validation of our model by reproduction of Venus IR results \citep{NRC2010}. The spacecraft is on a circular orbit at 0.7 au from the Sun. The camera's FOV = 11~$({}^\circ)^2$, 420 sec integration time, 60 sec slew time (covering 1,980~$({}^\circ)^2$/24 hours); threshold magnitude = 21.5.
\label{fig:8}}
\end{figure}

We found that using our simulation approach, the Venus IR mission required $\sim$7.8 yrs to find 90\% of H=22~mag EGs as in Fig.~\ref{fig:8}, which is within 10--20\% of the 7.5 yrs in \citep{NRC2010}.  A small discrepancy is expected because there may be differences in the modeling assumptions relative to \citep{NRC2010}.  However, overall, these results confirm our major assumptions and provide confidence that our general simulation approach is correct.

\subsection{Case 2: Survey time versus sky cover/camera FOV}
\label{sec:sky-cov-FOV}

We next studied the importance of the synthetic tracking camera FOV size in its effectiveness to detect 90\% of 140~m EGs.  The goal is to identify the optimal FOV size and to understand where our baseline CubeSat scanning strategy with $\sim$1,500 $({}^\circ)^2$/day is relative to these trades. 

Unlike conventional CCD-based exposures (which cannot trade integration time and sensitivity), with synthetic tracking we have the flexibility to trade the limiting magnitude for the sky coverage.  Doubling the integration time results in the SNR improving by $2^{1/2}$, the detection distance increasing by $2^{1/4}$, and the volume of space covered increasing by $2^{3/4}$; the sky coverage rate (in $({}^\circ)^2$/day), however, drops by a factor of 2.  With only one telescope, shorter integration times are preferred because they enable us to cover the sky more quickly, despite the reduction in SNR and sky volume coverage. It takes a finite amount of time to slew and settle the telescope, so that the telescope may capture the next area of sky.  For a given slew time, it is relatively easy to calculate the optimum exposure time, which is approximaterly three times the slew time when the goal is to maximize the number of objects viewed per hour.  However, when searching for EGs, maximizing the coverage per hour is not an appropriate objective. This is because the population of available objects changes slowly, on timescales of {\bf $\sim2$} months.  Therefore, there is a trade-off between the sensitivity (which improves with a smaller sky FOV) and the frequency of scanning the sky (which improves with a larger sky FOV).  
 
\begin{figure}[t]
%\epsscale{0.5}
%\plotone{fig9.eps}
\centering\includegraphics[width=7.0cm]{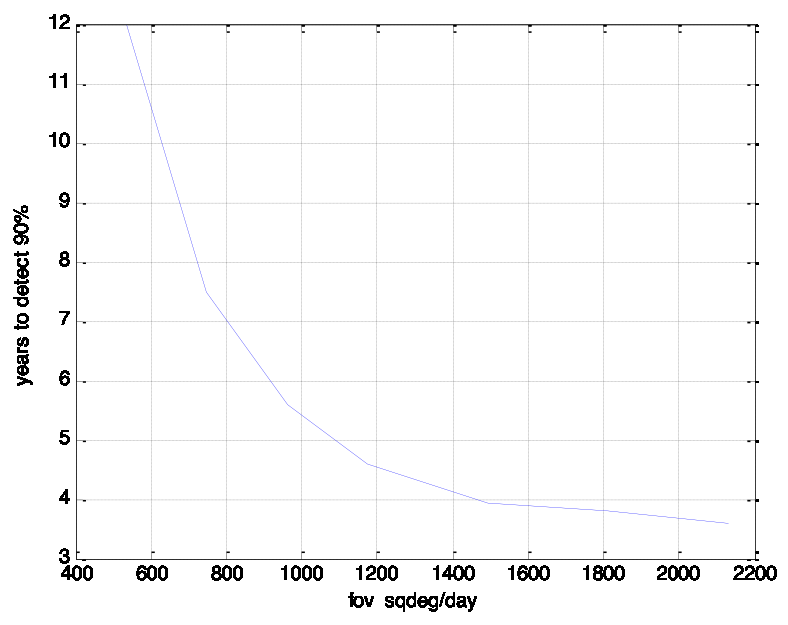}
\caption{
Time to find 90\% of $\rm H=22$~mag EGs detected as a function of FOV size and daily sky coverage for ideal constellation with six CubeSats at 0.85~au with variable FOVs, 800 sec integration, 10 sec slews and 20.49~mag limit when the telescope was at 1~AU.
\label{fig:9}}
\end{figure}

To evaluate the impact of the FOV on performance, we chose the constellation with six CubeSats at 0.85 au and simulated a range of FOVs between 0.25 and 25.0~$({}^\circ)^2$ (i.e., 600--2,400~$({}^\circ)^2$ in 24 hours), as shown in Fig.~\ref{fig:9} (assuming all had a 20.49  magnitude limit at 1~au). We note that for FOVs~$<3 ({}^\circ)^2$ the time to complete the survey exceeded 20 years, but the exact time was not solved; thus, these results are not shown in Fig.~\ref{fig:9}. The point where there are diminishing returns with increasing the FOV (or the knee in these plots) is around 14~$({}^\circ)^2$ (i.e., 1,500~$({}^\circ)^2$/day), where for greater FOVs, the time to detect 90\% of EGs was not affected significantly, but for smaller FOVs the detection time grows exponentially. Figure~\ref{fig:9} depicts the effectiveness of an EG search campaign as a function of the FOV of a telescope.  It shows that for a search for 90\% of EGs with H $\leq$ 22 with a facility that has a limiting magnitude of $\sim20.5$~mag, a scanning rate of more than $\sim1,500~({}^\circ)^2$/24 hrs is an inefficient use of resources.  The PanSTARRs telescope operating 10 hrs per night would survey 1,400 $({}^\circ)^2$ every observing night.  From these results we can conclude that increasing the sky coverage from 1,400~$({}^\circ)^2$ to 20,000~$({}^\circ)^2$ per night with more ground-based observatories would not significantly reduce the time to find 90\% of 140~m EGs.  A space mission in an Earth orbit duplicates the search volume of the ground-based observatories. 

\subsection{Case 3: Design of interplanetary CubeSat constellation}

Next, we investigate CubeSat constellation architectures to accomplish the mission.  The simulation environment described above was used to study different numbers of spacecraft in the constellation at various heliocentric ranges.  For the multi-spacecraft architectures, we assumed that the spacecraft were equally distributed in solar orbits. The telescope magnitude limit was assumed to be 20.49~mag at 1 AU, and zodi effects considered with FOV=14~$({}^\circ)^2$ (see Appendix~\ref{sec:cubesat}). The telescope is assumed to integrate for 800 sec and then the spacecraft slews for 10 sec (to model the analogous approach of a 400 sec observation and 5 sec slew time repeated twice between different pointings, which covers $\sim$1,500~$({}^\circ)^2$/24~hrs.  

\begin{figure}[!ht]
%\epsscale{0.45}
%\plotone{fig11.eps}
\centering\includegraphics[width=7.0cm]{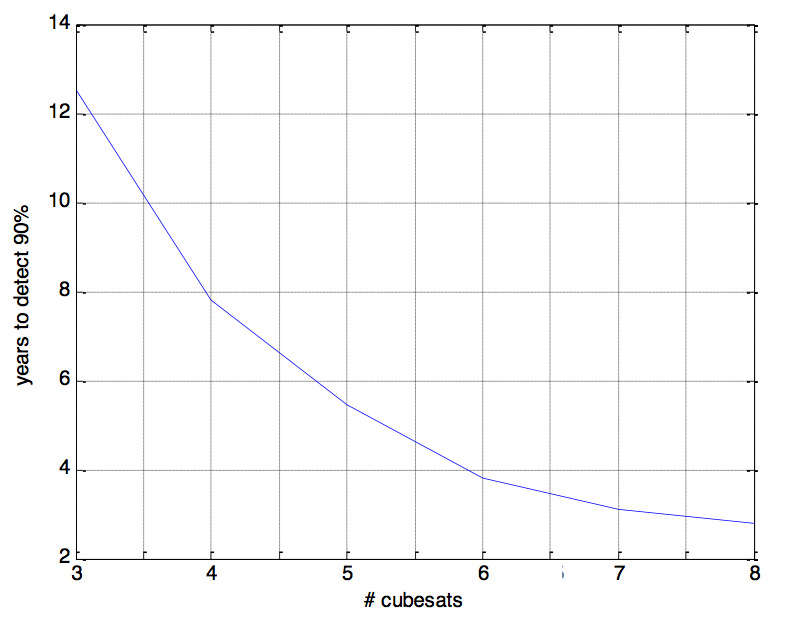}
\caption{Time to detect 90\% of H=22 mag EGs assuming FOV = 14$({}^\circ)^2$, 800 sec integration, 10~sec slews, covering $\sim$1,500~$({}^\circ)^2$ per 24 hours, with slewing model ``on''. The figure shows sensitivity to the number of CubeSats for constellations 0.85~au from the Sun.  In both cases, the CubeSats are distributed equally around the Sun.
\label{fig:11}}
\end{figure}

\begin{figure}[!ht]
%\epsscale{0.50}
%\plotone{fig13.eps}
\centering\includegraphics[width=7.0cm]{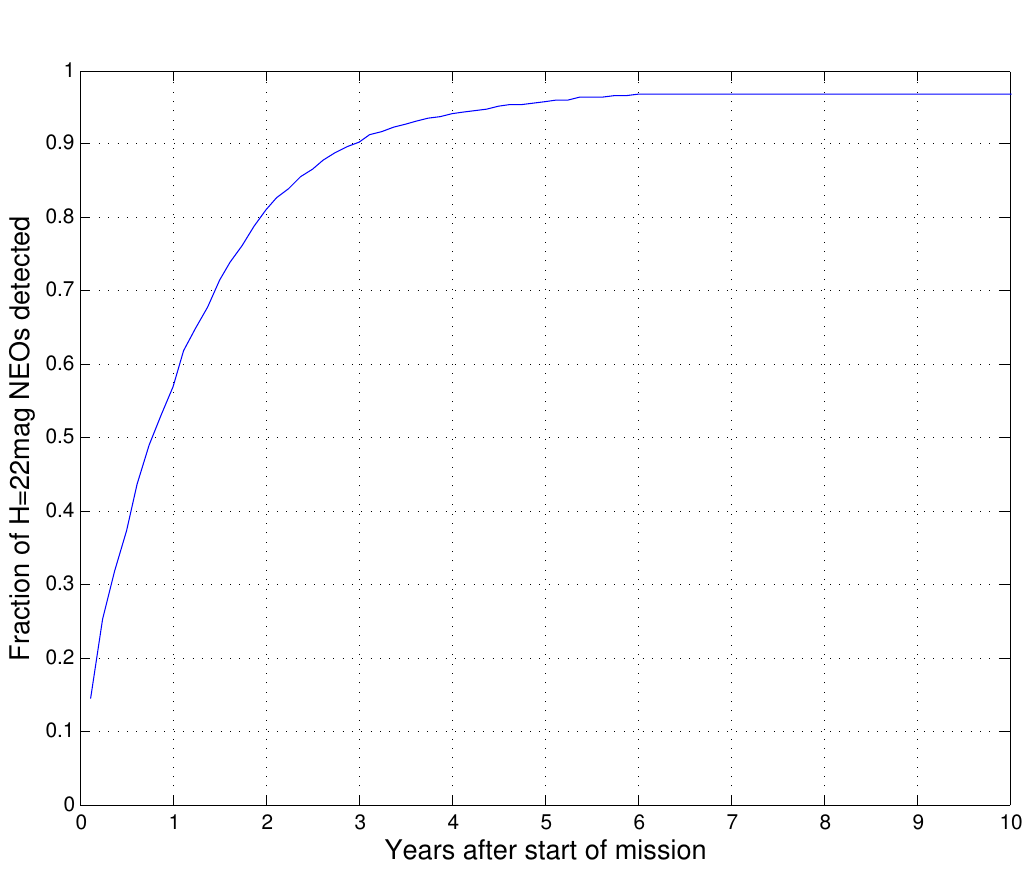}
\vskip -10pt 
\caption{
Fraction of H=22 mag EGs detected for constellations of CubeSats from Fig.~\ref{fig:11} with FOV = 14~$({}^\circ)^2$, 800 sec integration, 10~sec slews, covering $\sim$1,500~$({}^\circ)^2$/24~hrs, with slewing model active.  
\label{fig:13}}
\end{figure}

\begin{figure}[!h]
%\epsscale{0.45}
%\plotone{fig12a.eps}
%\plotone{fig12b.eps}
\centering\includegraphics[width=7.5cm]{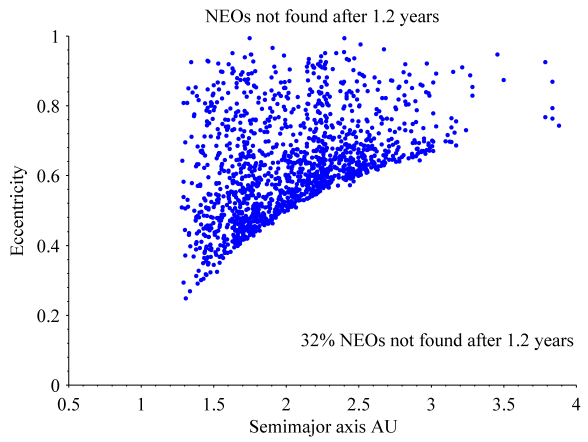}
\centering\includegraphics[width=7.5cm]{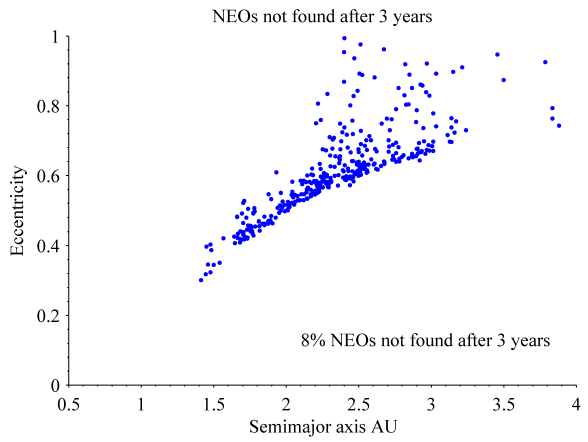}
\caption{Orbital properties of EGs that were not found after 1.2 years (left) and 3.0 years (right) with representative 
six-CubeSat constellations.
\label{fig:12}}
\end{figure}

We performed simulations for different constellation sizes at various heliocentric ranges from  0.7 to 1.1 AU. The results showing the time to detect 90\% of $\rm H=22$~mag EGs for an idealized constellation are in Fig.~\ref{fig:11}. There is a broad range of semi-major axis between 0.85 and 1.0~au where the time to find 90\% of EGs is minimized.  The minimum occurs due to three competing effects. As the CubeSat constellation gets closer to the Sun, the sky background increases which leads to reduced sensitivity and increased survey times. On the other hand, a larger and larger percentage of the EGs will be detected at a large phase angle, again reducing sensitivity when semi-major axis exceeds $\sim$1~au. Choice of smaller semi-major axis allows to minimize detection time because the orbits have shorter periods, covering the sky faster, and orbits that are closer to the perigees of the EGs. These competing effects produce the broad minimum between 0.85~au and 1.0~au. Looking at Fig.~\ref{fig:11}, we observe that addition of a new CubeSat to the constellation of $N$ spacecraft generally reduces the total time needed to complete the survey, where this reduction is most significant for small numbers of CubeSats. However, with each additional spacecraft, the additive search performance gain is reduced, with very little gain after $N \geq6$.
Fig.~\ref{fig:13} shows the time to find 90\% of H=22 mag EGs with a constellation of six CubeSats placed on orbits with a semi-major axis of $\sim 0.85$~au. 

We also note that the mission time never falls below 2 years, even for constellations with up to $N=10$. This is due to the fact that with six uniformly-distributed spacecraft in solar orbit it takes nearly 2 years to find almost 90\% of all the EGs with magnitude less than H=22. The remaining ``undiscovered'' EGs have semi-major axis ranging from 2--3~AU, spending most of their orbits beyond the accessibility of the CubeSats placed on solar orbit with semi-major axis of 0.85--1.0~AU, as in Fig.~\ref{fig:12}.  Capturing these EGs is simply a waiting game until they approach their periapsis and are close enough to a spacecraft to be detected. 

\subsection{Case 4: Detection of smaller EGs}

\begin{figure}[ht]
%\epsscale{0.55}
%\plotone{fig14.eps}
\includegraphics[width=8.0cm]{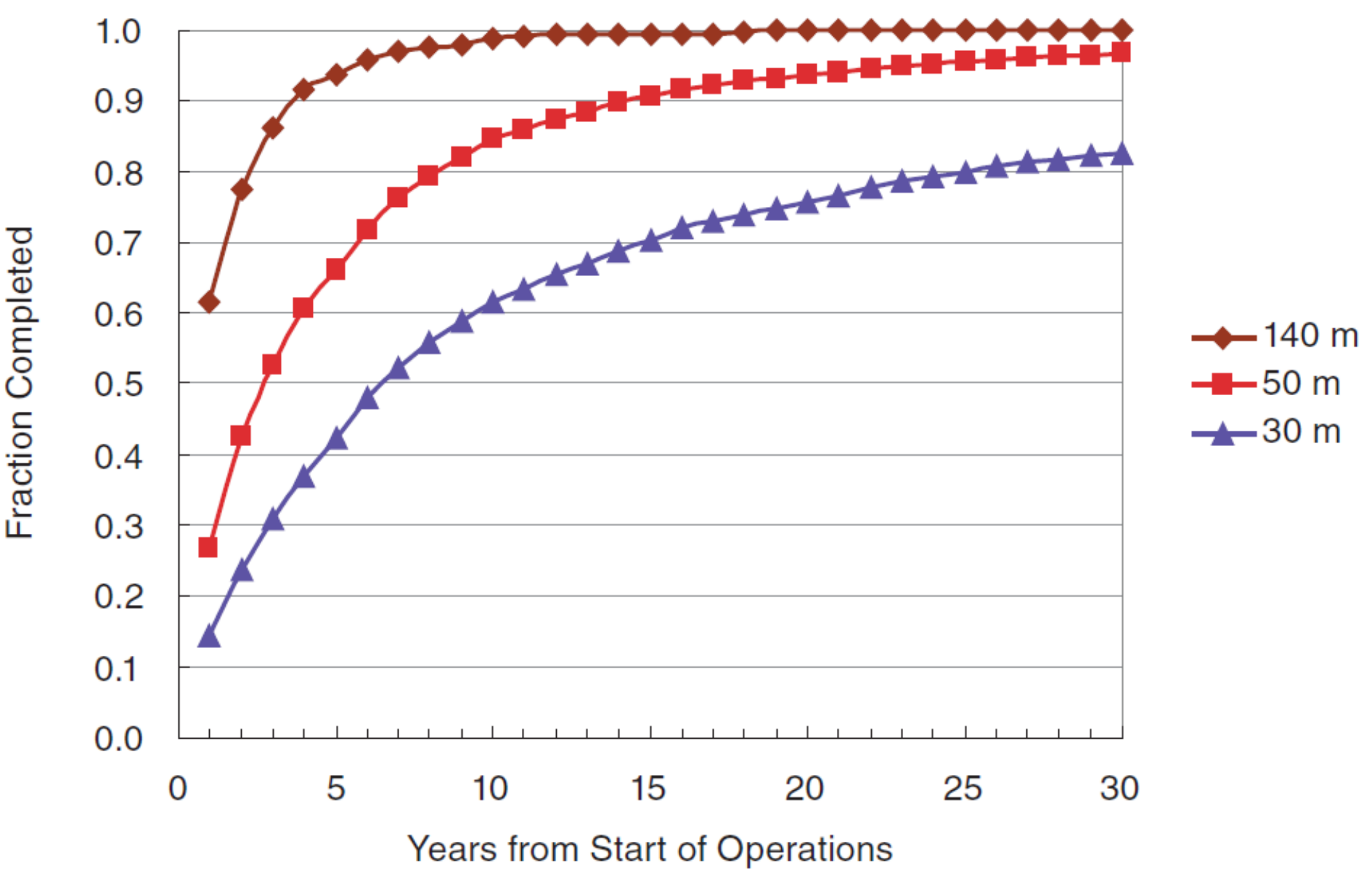}
\caption{Years to completion for 0.5~m IR telescope in a Venus-like orbit and a dedicated LSST [Fig. 3.10 from \citep{NRC2010}].
\label{fig:14}}
\end{figure}

The 2010 NRC report concluded that a survey to find 90\% of 140~m EGs would require a space mission with a lifetime of $\sim$8--10 years.  In addition to surveying for large EGs, the report also emphasized the need to look for smaller EGs, with sizes down to 30~m, for planetary protection purposes.  Fig.~\ref{fig:14} is from the NRC report showing the effectiveness of an EG search campaign with a combined IR telescope in a Venus-like orbit and a dedicated LSST.  If completely dedicated to EG search, LSST would be a very powerful facility. If we ignore the loss of sensitivity from streaked images, LSST would be able to detect 25~mag objects.  However, the combination of these two facilities represents a relatively large investment, and even then, detection of 90\% of 50~m EGs would take $\sim$14 years. The NRC report ended by saying that detection of these smaller EGs is extremely challenging and expensive. We note that a typical operating cost of a major observatory is $\sim$10\% of it's construction cost per year. In the case of  LSST, whose total construction cost is estimated to be $\sim$\$700M\footnote{To that extent, see the relevant discussion of the costs involved at {\tt https://www.nsf.gov/news/news$\_$summ.jsp?cntn$\_$id=124899}}),  operating it even for a few years while 100\% devoted to NEO search would be much more expensive than a constellation of Cubesats.  
 
We re-examined this conclusion with a constellation of CubeSats with synthetic tracking cameras. Specifically, given the higher sensitivity enabled by synthetic tracking and high sky coverage rates at affordable costs provided by CubeSat platforms, we were interested in the potential for the proposed architecture to detect 90\% of 50~m EGs in less than 10~years. 

The 50~m EGs are almost a factor of 10 dimmer than the 140~m EGs, with H =24.2~mag. If the observatory is at 0.7~AU, a 140~m EG (H=22~mag) could be detected at ranges up to 0.4~au away, if the detection magnitude limit was 20.5~mag. However, an EG with H=24.2~mag ($\sim$50~m in size) would have to be at a range of no greater than 0.20~au from the telescope and at opposition to be detected. At any one time, the volume of the search space is eight times smaller. The search for these smaller EGs would require a more capable constellation than the strawman constellation we described for H=22~mag EGs.  We also make some assumptions as to what type of detectors and small telescopes would be available in the near future.  We pick one constellation as an example to illustrate the power of this approach realizing that optimizing the constellation design requires more work.

We assume that $\tt (8K\times8K)$ detectors will become available, with other performance parameters similar to current smaller-format sCMOS detectors.  The assumptions for this much more capable camera/constellation are listed in the Table~\ref{tab:adv-st-camera}. This constellation would detect 90\% of 50~m EGs (H=24.2~mag) more than once in approximately 4.5 years, and compares favorably to all the concepts examined in the 2010 NRC report.

\begin{table}[h]
\begin{center}
\begin{scriptsize}
\caption{Parameters for an advanced synthetic tracking camera.
\label{tab:adv-st-camera}}
\vskip 5pt
\begin{tabular}{|r|c|c|}
\hline\hline
Parameters & Value & Units \\\hline
Telescope Diameter & 20 & cm \\\hline
Image& 2X & diff limit  \\\hline
Pixel size &    1.6&    arcsec\\\hline
Effective background    &2.8&   arcsec\\\hline
Magnitude limit &22.15& mag\\\hline
Integration time        &800    &s\\\hline
Number of satellites &  {\bf 9}&\\\hline        
EG size &50&    m\\\hline
H magnitude     &24.24 & mag\\\hline
\hline
\end{tabular}
\end{scriptsize}
\end{center}
\end{table}

\begin{figure}[!ht]
%\epsscale{0.45}
%\plotone{fig15.eps}
\centering\includegraphics[width=7.0cm]{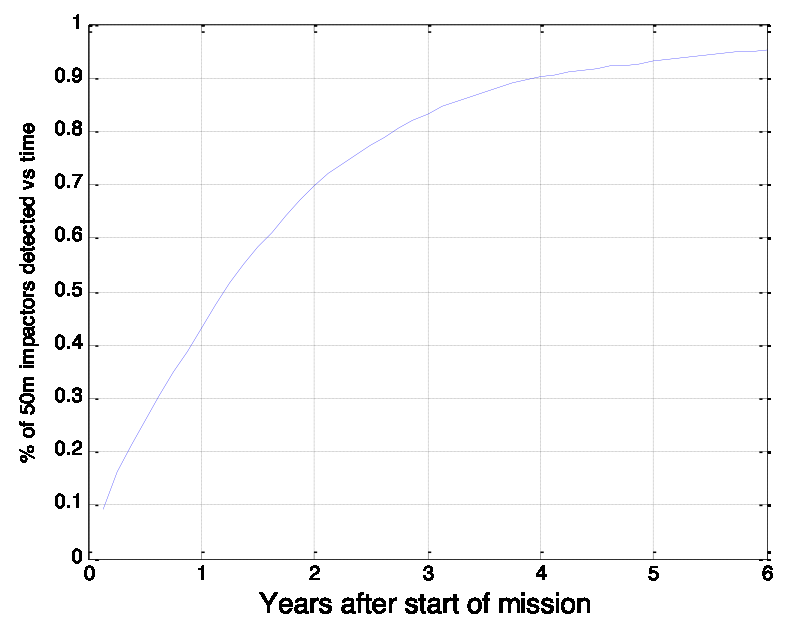}
\caption{Time to detect 90\% of 50~m EGs with the constellation of SmallSats from Table~\ref{tab:adv-st-camera}.
\label{fig:15}}
\end{figure}

In solar orbits at 0.85 AU, and a solar avoidance angle of $65^\circ$, a constellation of nineCubeSats with 20 cm apertures will detect 90\% of 50 m EGs in $\sim 4.0$~years, as shown in Fig.~\ref{fig:15}.

This example illustrates the potential for a constellation of CubeSats to conduct a 90\%-complete search for 50 m EG that, according to the NRC report, would otherwise take a long time at a prohibitively high cost. These examples are not yet optimized to identify the configuration with minimal detection time or cost. Nevertheless, they show synthetic tracking with CubeSats may be the only affordable architecture to search for EGs significantly smaller than 140~m.

The results above demonstrate the advantages of the proposed approach based on a constellation of CubeSats with synthetic tracking to search for EGs with various sizes. Such a constellation could conduct a 90\% complete survey in significantly less time and lower cost relative to what was indicated in the NRC report. A more detailed and thorough examination of CubeSat constellations aiming at detecting 90\% of $\sim$35--100~m EGs will be the topic of a future paper.

\section{Discussion and summary}
\label{sec:discussion}

The time required to detect 90\% of EGs with H~$\leq$ 22~mag depends on a number of factors, including limiting magnitude sensitivity, size of the FOV, and number and configuration of spacecraft. Our 10~cm CubeSat-based camera and ground-based telescopes such as ZTF\footnote{The Zwicky Transient Facility (ZTF) is a new time-domain survey that will have first light at Palomar Observatory in 2017. For details, please visit: {http://www.ptf.caltech.edu/ztf}}, CSS, and ATLAS have a limiting magnitude of $\sim$20.5 mag. Larger telescopes, such as PanSTARRs and LSST with smaller pixels (\arcsec/pixel) are more sensitive. Space-based 50~cm IR telescopes are also more sensitive ($\sim$21.5 mag) relative to similar ground-based visible cameras. Simulations conducted for \citep{NRC2010} showed the importance of the distribution of the observatories. For ground-based observatories, a distribution in geographic latitude helps with sky coverage. The NRC report also hinted at the importance of distributing observatories around the solar system. The most capable combination of observatories examined in \citep{NRC2010} was LSST and an IR telescope in a Venus orbit. A Venus-like orbit is advantageous relative to one on Earth or in an Earth orbit because this type of telescope would essentially duplicate much of what LSST would detect.

Our survey simulations in Sec.~\ref{sec:sky-cov-FOV} show that sky coverage beyond $\sim$1,400~$({}^\circ)^2$ per 24 hrs would not significantly decrease the time needed to find 90\% of H=22 mag EGs. It should be noted that PanSTARRs with its 7~$({}^\circ)^2$ FOV and spending $4\times(30~{\rm sec}+15~{\rm sec})$ per FOV would cover $\sim$1,400~$({}^\circ)^2$ in 10 hrs. Adding more ground-based telescopes without increasing sensitivity would not shorten the time to conduct a 90\% complete survey.

Earlier in the introduction we mentioned  that reaching high astrometric precision in measuring the orbital parameters of EGs is as important as detecting them. For planetary protection purposes, the goal of finding 90\% of H=22 mag EGs is not sufficient if we do not have the data to determine weather or not a newly discovered asteroid will impact the Earth or not. If the astrometric accuracy is 0.2\arcsec, and the EG is observed a half dozen times on its first pass, for the observational data arc of 13 days the orbit derived from those measurements is rather poor. In fact, it is so poor that if the EG were to revisit $\sim$4--5 years later, we would not know where to point the telescope within $\approx10^\circ$.

Currently a high accuracy orbit will require one of two circumstances:  One is where several observations during one apparition are augmented by radar observations. The other is where observations at multiple apparitions are possible. Our simulations of a nominal six-SmallSat constellation show that, on average, each EG is detected  approximately ten times by several different SmallSats over a $\sim$6~yr period.  Our simulations also showed that $\sim 78$\% of the EGs would be detected 4 or more epochs over $a >120^\circ$ orbital arc. These $\sim$4+ observations, each with $\sim$0.5\arcsec~astrometry per observation, may be sufficient to measure the EG orbit to measure the MOID with Earth's orbit to a few times the Earth's diameter. A more detailed examination of this issue will be addressed elsewhere.
  
Here we look at a constellation of SmallSats in solar orbit capable of detecting 90\% of EGs one or several times during the entire observing campaign.  We compared results for various SmallSat constellations to the results of the 2010 NRC report. We note, that if only a single detection of an asteroid is made, that is far from what is needed to determine if that asteroid would eventually hit the Earth. Ideally, a survey would not only detect the objects, but detect them enough times to measure an accurate orbit. If accurate orbit measurements of a significant fraction of EGs is not possible, then it is preferable to have enough measurements of EGs that those measurements can be linked with the measurements of another survey.  These topics are beyond the scope of this paper and will be addressed in a subsequent paper currently in preparation.

We briefly address the topic of linking observations from a constellation of SmallSats. To study this problem, in one of the simulations, we adopted a different scan pattern, one designed to enhance ``linked'' observations. Instead of spending 800 sec then moving to an adjacent FOV, we split the 800 sec into two 400 sec blocks 2 hrs apart. The entire 800 sec was treated as a single data cube, as discussed in Sec.~\ref{sec:3.2}. This resulted in a velocity measurement that was $\sim$0.5~\arcsec/(2~hr). That same area was again scanned 2 days later again with 800~sec total integration time. This cadence would be repeated $\sim$25 days later when the whole sky was rescanned. The 2nd epoch has to be $\sim$25 days later in order for the two sets of measurements to be connected with low false-positive and low false-negative connection. The original scan pattern 810 sec for a $14.1~({}^\circ)^2$ FOV, would scan $\sim 3\pi$ steradian every $\sim$20 days. This doubled scan pattern would have a natural ``repeat'' cycle time of $\sim$40 days. In order to reduce this time, the ``linked'' scan pattern had a slightly larger Sun-avoidance angle, but more importantly it avoids scanning the ecliptic poles ($\pm45^\circ$ ecliptic latitude). 

This nominal constellation of six SmallSats would provide linked observations of 79\% of H=22~mag EGs in six years and 95\% of EGs would be detected at least once. This cadence of observations, called a ``bottom up'' approach, enables the linking observations of the same object observed at different times by different satellites. The other approach is a ``top down'' approach. With astrometric accuracy $<1$\arcsec, observation of the same asteroid four times when the orbital arc is $\gtrsim 120^\circ$ results in a very accurate orbit. If three of the observations were of one asteroid and the 4th of a 2nd asteroid, the fit residuals would be large.

The low cost of an interplanetary CubeSat presents the opportunity of a mission architecture consisting of launching multiple spacecraft in solar orbit. With low anticipated cost, this option is not only feasible, it is also highly attractive. In fact, a mission relying on multiple spacecraft allows for added mission redundancy, effective sky coverage, and a shorter period to complete the EG search and provides chances for frequent technology upgrades. With preliminary cost estimates for interplanetary SmallSats being so low, a further cost reduction would come from the fact that additional CubeSats would cost only a fraction of the original cost (50\% or less due to recurring engineering costs). Therefore, the cost of launching six to ten interplanetary SmallSats is still expected to be an order of magnitude lower than that for the missions in the 2010 NRC report. 

Significant improvements are expected from optimizing the constellation design, scanning strategies, FOVs, and anticipated SmallSat lifetimes. A rigorous analysis of the optimal number of spacecraft to mitigate risks and ensure a high probability of mission success in the desired time for an acceptable cost is beyond the scope of this paper; it will be presented elsewhere.

The mass-produced space-qualified hardware used in small satellites dramatically reduces the cost of a space observatory making a constellation of these telescopes not only affordable, but also significantly lower in cost relative to conventional medium-sized space telescopes such as those in the NRC report. Furthermore, EGs much smaller than H $\leq$ 22 mag (or $\rm \sim140$~m in size) can still cause major damage when they impact the Earth. We emphasize that the only affordable way to survey 90\% of 70~m or 50~m EGs would be with the synthetic tracking multiple SmallSat architecture. We will investigate the relevant mission design and architecture in a subsequent publication.  

In conclusion, we observe that by combining synthetic tracking and CubeSat technologies, compared to all survey architectures and methods proposed previously, we are fundamentally ``playing in a different ball park''. This new paradigm is both much less expensive and significantly more capable of finding not just 140~m EGs in much less time, but also 90\% of smaller EGs with sizes down to 50 m. 
We note that even a factor of two increase in our cost estimates would still be a fraction of the cost of the missions/facilities mentioned in \citep{NRC2010}.  
Clearly, more mission design work is needed. Therefore, the potential of a constellation-based architecture presented here will be explored further.

\begin{acknowledgements}
We thank Paul Chodas, Steve Chesley, T. Joseph W. Lazio and Robert Preston of JPL for their interest and many useful discussions during the work and preparation of this manuscript. It is our pleasure to thank Prof. Edward (Ned) L. Write of UCLA for valuable comments and suggestions. We also thank Mikael Granvik for valuable comments on the manuscript and also for allowing us to use the NEO distribution model from \citep{Granvik-etal:2016}. The work described here was carried out at the Jet Propulsion Laboratory, California Institute of Technology, under a contract with the National Aeronautics and Space Administration. 
\end{acknowledgements}

%\bibliography{cubesat-NEO}

\begin{thebibliography}{23}
\expandafter\ifx\csname natexlab\endcsname\relax\def\natexlab#1{#1}\fi

\bibitem[{{Bernstein} {et~al.}(2004){Bernstein}, {Trilling}, {Allen}, {Brown},
  {Holman}, \& {Malhotra}}]{Bernstein-etal:2004AJ}
{Bernstein}, G.~M., {Trilling}, D.~E., {Allen}, R.~L., {et~al.} 2004, \aj, 128,
  1364

\bibitem[{{Boslough} \& {Crawford}(1997)}]{Boslough-Crawford:1997}
{Boslough}, M. B.~E. \& {Crawford}, D.~A. 1997, Annals of the New York Academy
  of Sciences, 822, 236

\bibitem[{{Boslough} \& {Crawford}(2008)}]{Boslough-Crawford:2008}
{Boslough}, M. B.~E. \& {Crawford}, D.~A. 2008, International Journal of Impact
  Engineering, 35, 1441 , hypervelocity Impact Proceedings of the 2007
  Symposium HVIS 2007

\bibitem[{{Chyba} {et~al.}(1993){Chyba}, {Thomas}, \&
  {Zahnle}}]{Chyba-etal:1993}
{Chyba}, C.~F., {Thomas}, P.~J., \& {Zahnle}, K.~J. 1993, \nat, 361, 40

\bibitem[{{Duncan} {et~al.}(2014){Duncan}, {Smith}, \&
  {Aguirre}}]{Duncan-etal:2014}
{Duncan}, C.~B., {Smith}, A.~E., \& {Aguirre}, F.~H. 2014, in Proceedings of
  the AIAA/USU Conference on Small Satellites, {\rm paper: SSC14-IX-3}, {\tt
  http://digitalcommons.usu.edu/smallsat/2014/AdvTechComm/3}, Logan, Utah, USA

\bibitem[{{Fazio} {et~al.}(2014){Fazio}, {MacNeal}, \&
  {Sheldon}}]{Sheldon-etal:2014}
{Fazio}, M., {MacNeal}, K., \& {Sheldon}, D. 2014, in Electronics Technology
  Workshop, NASA Electronic Parts and Packaging Program (NEPP)

\bibitem[{Granvik {et~al.}(2016)Granvik, Morbidelli, Jedicke, Bolin, Bottke,
  Beshore, Vokrouhlick{\'y}, Delb{\`o}, \& Michel}]{Granvik-etal:2016}
Granvik, M., Morbidelli, A., Jedicke, R., {et~al.} 2016, Nature, 530, 303

\bibitem[{{Granvik} \& {Muinonen}(2005)}]{Granvik-Muinonen:2005}
{Granvik}, M. \& {Muinonen}, K. 2005, \icarus, 179, 109

\bibitem[{{Granvik} \& {Muinonen}(2008)}]{Granvik-Muinonen:2008}
{Granvik}, M. \& {Muinonen}, K. 2008, \icarus, 198, 130

\bibitem[{{Granvik} {et~al.}(2007){Granvik}, {Muinonen}, {Jones},
  {Bhattacharya}, {Delbo}, {Saba}, {Cellino}, {Tedesco}, {Davis}, \&
  {Meadows}}]{Granvik-etal:2007}
{Granvik}, M., {Muinonen}, K., {Jones}, L., {et~al.} 2007, \icarus, 192, 475

\bibitem[{{Gural} {et~al.}(2005){Gural}, {Larsen}, \&
  {Gleason}}]{Gural-etal:2005AJ}
{Gural}, P.~S., {Larsen}, J.~A., \& {Gleason}, A.~E. 2005, \aj, 130, 1951

\bibitem[{{Harris}(2009)}]{Harris:2009}
{Harris}, A.~W. 2009, {The NEO population}, impact risk, progress of current
  surveys, and prospects for future surveys, presentation to the Survey,
  Detection panel of the nrc committee to review near-earth object surveys and
  hazard mitigation strategies, {Jan. 28-30, 2009}, Space Science Institute

\bibitem[{{Harris}(2011)}]{Harris:2011}
{Harris}, A.~W. 2011, The NEO population, presentation at the target neo
  workshop, george washington university, feb. 22, 2011, {\tt
  http://targetneo.jhuapl.edu/pdfs/sessions/targetneo-session2-harris.pdf},
  Space Science Institute, Consultant to NASA/JPL NEO Program Office

\bibitem[{{Leinert} {et~al.}(1982){Leinert}, {Richter}, {Pitz}, \&
  {Hanner}}]{Leinert-etal:1982}
{Leinert}, C., {Richter}, I., {Pitz}, E., \& {Hanner}, M. 1982, \aap, 110, 355

\bibitem[{Marrese-Reading {et~al.}(2010)Marrese-Reading, Ziemer, Scharf, Tomas
  J. Martin-Mur, Mueller, , \& Wirz}]{Marrese-Reading-etal:2010}
Marrese-Reading, C.~M., Ziemer, J.~K., Scharf, D.~P., {et~al.} 2010, in
  Proceedings of the IEEE Aerospace Conference, Big Sky, MT

\bibitem[{{Mueller} {et~al.}(2010){Mueller}, {Hofer}, \&
  {Ziemer}}]{Mueller-etal:2010}
{Mueller}, J., {Hofer}, R., \& {Ziemer}, J. 2010, in Proceedings of the 57th
  Joint Army-Navy-NASA-Air Force (JANNAF) Propulsion Meeting, Colorado Springs,
  CO

\bibitem[{NRC(2010)}]{NRC2010}
NRC. 2010, {Defending Planet Earth: Near-Earth Object Surveys and Hazard
  Mitigation Strategies}, Tech. rep., National Research Council

\bibitem[{{Roach} \& {Gordon}(1973)}]{Roach-Gordon:1973}
{Roach}, F.~E. \& {Gordon}, J.~L. 1973, {The Light of the Night Sky},
  Geophysics and Astrophysics Monographs, (Book 4) (Springer, (December 1,
  1973))

\bibitem[{{Shao} {et~al.}(2014){Shao}, {Nemati}, {Zhai}, {Turyshev}, {Sandhu},
  {Hallinan}, \& {Harding}}]{Shao-etal:2014}
{Shao}, M., {Nemati}, B., {Zhai}, C., {et~al.} 2014, \apj, 782, 1

\bibitem[{{Spangelo} {et~al.}(2015){Spangelo}, {Dalle}, \&
  {Longmier}}]{Spangelo_JSR2014}
{Spangelo}, S., {Dalle}, D., \& {Longmier}, B. 2015, AIAA/AAS Astrodynamics
  Specialist Conference (San Diego, CA, 2015),
  http://dx.doi.org/10.2514/6.2014-4125

\bibitem[{{Tanga} \& {Mignard}(2012)}]{Tanga-Mignard:2012}
{Tanga}, P. \& {Mignard}, F. 2012, Planetary and Space Science, 73

\bibitem[{{Zarifian} {et~al.}(December 17, 2014)}]{Zarifian-etal:2014}
{Zarifian}, P. {et~al.} December 17, 2014, Asteroid Detection with CubeSats
  using synthetic tracking, {JPL TeamXc} report \#1555 on the study 2014-12,
  JPL

\bibitem[{{Zhai} {et~al.}(2014){Zhai}, {Shao}, {Nemati}, {Werne}, {Zhou},
  {Turyshev}, {Sandhu}, {Hallinan}, \& {Harding}}]{Zhai-etal:2014}
{Zhai}, C., {Shao}, M., {Nemati}, B., {et~al.} 2014, \apj, 792, 60

\end{thebibliography}

\begin{appendix}
\section{Search for potentially hazardous NEOs with the Synthetic Tracking Technique}
\label{sec:syn-track}

As NEOs are moving against the background stars, their resulting images when using a conventional CCD are streaked, resulting in degradation  in sensitivity. The NRC report included a simulation of a hypothetical 2~m space telescope conducting a NEO search. The optimal integration time for this instrument was only 8 sec with 30 sec of slewing between adjacent FOVs
[S. Chesley of JPL (2015), private communication]. Longer CCD exposures do not improve sensitivity because of image streaking. The technique of co-adding multiple frames of data to stack the images of faint object was first developed and demonstrated for trans-Neptunian objects (TNO) \citep{Bernstein-etal:2004AJ}. The images were taken over periods of months. \cite{Gural-etal:2005AJ} used the same technique, also called multi-hypothesis matched filter (MHMF), for processing on asteroids with images taken over a period of minutes/hours. Synthetic tracking combines MHMF with faster cameras to detect smaller objects moving at higher angular velocities. This technique makes it possible to achieve a sensitivity down to a $\sim$20.5 mag object with 10 cm optics in an 800 sec observation, thus offering significant improvements in the sensitivity of detection of NEOs. Below we discuss this technique as it applies for a CubeSat-based version of a synthetic tracking camera.  

\subsection{Improving sensitivity, SNR, and astrometry with synthetic tracking} 
\label{sec:syn-track_snr}

Traditional approaches to discovering NEOs relies on CCD exposures of $\sim$30~sec. Typically, CCDs require $\sim$10 sec read-out time at low noise ($\sim$3e$^-$). Although this approach is effective in detecting slowly moving NEOs, for faster objects it results in a streaked image on the CCD and leads to a significant trailing loss of sensitivity. Intuitively, trailing loss results from the fact that the streaked image distributes photons comprising its signal over a larger area on the CCD (compared to those received from a stationary object) yielding a reduced signal per unit area. There have been many studies on trailing losses.  \citet{Shao-etal:2014} quantify the trailing loss in SNR as a factor $w/(w+s)$, where $w$ is the width of a seeing-limited point-spread function (PSF) and $s$ is the length of the streak. The longer exposure leads to a longer streak length and results in a smaller SNR. 
 
\begin{figure}[ht]
%\epsscale{0.5}
%\plotone{fig1.eps}
\centering\includegraphics[width=7.0cm]{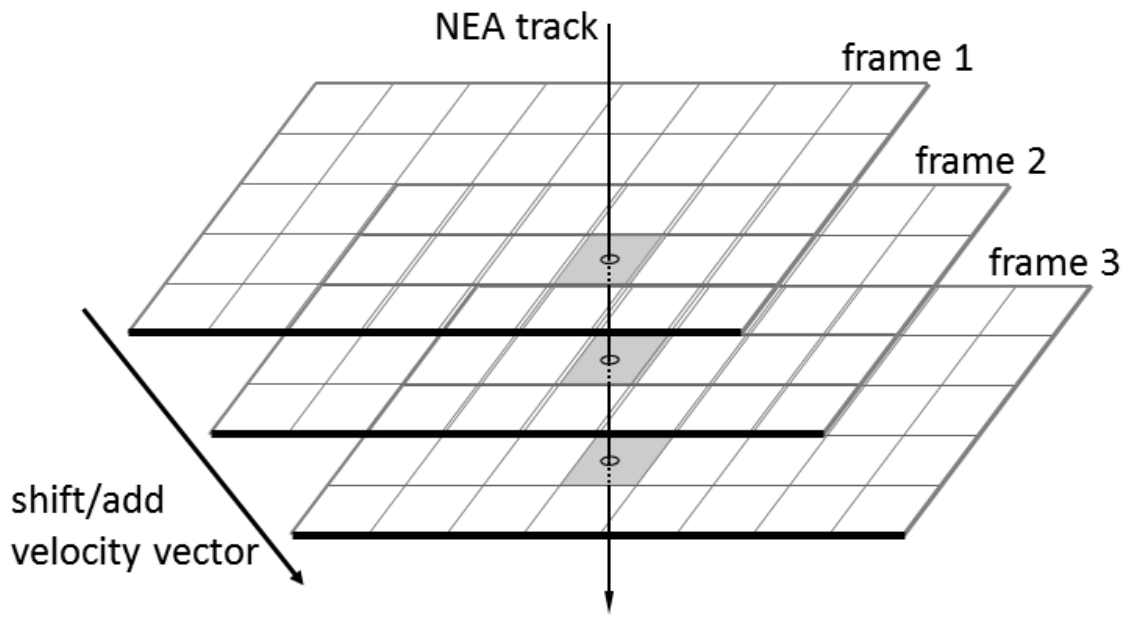}
\caption{Schematic showing the integration of frames by using synthetic tracking. Frames are displaced according to the velocity of a NEO so that it is at the same location in all the frames during the integration (adopted from \citep{Shao-etal:2014}).
\label{fig:1}}
\end{figure}

Compared to the conventional method of a single 30 sec exposure, synthetic tracking uses multiple short exposures: for example, 60 frames at 2 Hz over the same 30 sec interval.  For a small object, a single 0.5~sec exposure image is not sufficient to detect the object in one frame; instead, an addition of appropriately shifted image frames is needed to reconstruct the image of the object. Figure~\ref{fig:1} illustrates the shift/add technique. We shift each subsequent image by an assumed velocity vector.  If that assumed velocity is the actual NEO velocity, all the NEO photons will end up in the same pixels in the stacked image. However, for an unknown NEO with an unknown velocity, many different velocities must be tried to determine the true velocity.

NEOs are found by conducting a 4D search in our 3D data cube for $(x,y,v_x,v_y)$, which are positions and velocities of NEOs. This effort is computationally intensive \citep{Zhai-etal:2014}. For a ground-based facility, the search is done on a graphics-processing unit (GPU) with 2500 cores with a velocity grid of size $100\times100$, with velocity grid spacing of 1{\arcsec} per 30 seconds.  This ensures that the maximum velocity error when searching for NEOs is less than 0.5{\arcsec} in 30 sec, which means that the images are streaked by less than 0.5{\arcsec} along right ascension (RA) or declination (DEC), which is a negligible trailing loss for 1{\arcsec pixels}. A typical velocity-searching range covers $\pm40~{}^\circ$/day in both RA and DEC, which is adequate for most NEO detections. A maximal velocity of 40 ${}^\circ$/day is enough to cover over 99.9\% of all NEOs.  However, even faster-moving objects will be detected but result in streaked NEO images and therefore lower sensitivity.
 
\begin{figure}[ht]
%\epsscale{0.7}
%\plotone{fig2.eps}
\centering\includegraphics[width=9.0cm]{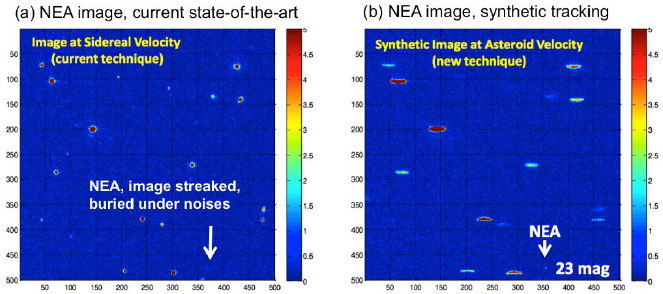}
\caption{Synthetic tracking images from integrating more than 500 frames taken at 17~Hz at sidereal velocity (a) and velocity of asteroid (b) (adapted from \citep{Shao-etal:2014}). Horizontal axis is the pixel number; vertical axis is the pixel number (left) and the signal intensity (right, in color).
\label{fig:2}}
\end{figure}

We have demonstrated the performance of the synthetic tracking technique towards improving the detection SNR by successfully detecting a faint object with an apparent magnitude of 23 (H$\sim$29.5 mag assuming the asteroid velocity is 10~km/s detected at 20 lunar distances) on the Palomar 5 m telescope \citep{Zhai-etal:2014}. The object was moving at $\sim$6 ${}^\circ$/day, covering $\sim$7{\arcsec} during the 30 sec observation time. Figure~\ref{fig:2} shows (a) the synthetically tracked images for tracking at the sidereal rate and (b) the asteroid from integrating more than 500 frames taken at 17 Hz by an EMCCD\footnote{Electron Multiplying Charge Coupled Device (EMCCD), see details at {http://www.emccd.com/what\_is\_emccd/}} with a negligible read noise when used with an EM gain of 200.  Image (a) would be the image detected by using the traditional 30 sec exposures. The asteroid image is a 7{\arcsec} streak with a surface brightness of a 25 mag star, with a sky background of $\sim$21 mag/({\arcsec})$^2$.  We detected this object, shown in (b), with an SNR$\sim$15. 
 
The detection above is a good example of using the synthetic tracking technique for detection of a previously unknown, small, fast-moving and otherwise undetectable faint object.  It also demonstrates the maturity and functionality of our software that is capable of removing detector artifacts, stars and galaxies, as well as identifying false positives. 

Synthetic tracking, in general, is much less sensitive to false positives compared to traditional ``tracklet'' identification of asteroids. With a data cube with $\sim$60 consecutive images, cosmic rays are easily removed. With single exposure NEO images, galaxies may mimic a slightly streaked image of a NEO. However, with synthetic tracking, the size of the object is minimized when correct velocity is identified. Therefore, galaxies are never mistaken for NEOs in synthetic tracking because the elongation of the image is minimum at zero velocity. In the ``tracklet'' approach to NEO detection, the image is ``thresholded to identify objects''. Objects that move linearly across $\sim$3--4 images taken over $\sim$10--30 minutes are identified as NEOs. Because each image is ``thresholded'' to identify objects, stars whose brightness is near the threshold will appear above the threshold in some images and not others. Stellar false positives are not a problem for synthetic tracking, first because we typically have many tens of images in a data cube, and second, at non-zero velocity, faint stars near the threshold will be streaked and be fainter than they would be if the images were co-added with zero velocity. 

Synthetic tracking also improves astrometry of NEOs. It accomplishes this in two ways. First, by mitigating the trailing loss, one achieves more precise measurements due to a higher SNR.  Second, it cancels a number of leading error sources that dominate traditional NEO searches, especially those from the ground \citep{Zhai-etal:2014}.  Thus, in CCD astrometry of a 2D point source, a template PSF\footnote{The point spread function (PSF) describes the response of an imaging system to a point source or point object.} is fitted to the CCD data.  In synthetic tracking astrometry, a moving template is fitted to the 3D data cube.  Using the images from the data cube, neither the asteroid nor the background stars are streaked.  Therefore, the image motion from the atmosphere and telescope tracking errors are now common between the NEO and background stars and, thus, cancel for relative astrometry. 

Observations from space are less affected from astrometric errors that are present for ground-based observations. Such an advantage allows for longer integration times yielding higher astrometric precision, which is why deploying a synthetic tracking technique on a space telescope is compelling.

\subsection{Moderate sensitivity from a small telescope}

In conventional ground-based NEO searches, it does not make sense to take a CCD exposure that exceeds $\sim$30~sec.  A NEO at a distance of 0.4 au moving 10~km/s would appear to move relative to background stars by 1{\arcsec} in 30~sec. Thus, for a ground-based telescope with 1{\arcsec} pixel, 30~sec is close to the optimal exposure time. Longer exposures would not only produce a streak, but they would also increase the background noise contribution without increasing the signal.  On the other hand, with synthetic tracking, we can observe for a much longer time, $T$, than 30~sec with increased SNR as $\sqrt{T}$.

The sensitivity of a synthetic tracking camera depends on a number of parameters: telescope diameter, pixel size and the total observation time, assuming the individual exposures are short enough that the motion of the NEO is less than 1 pixel.

\begin{table}[h!]
{\small
\begin{center}
\caption{Parameters used to estimate sensitivity of synthetic tracking camera. Total QE describes a combined QE, which includes the system's optical throughput  and detector QE; otherwise detector QE is used.
\label{tab:param}
}
\vskip 5pt
\begin{scriptsize}
\begin{tabular}{r|c|l}
\hline\hline
& Value  & Unit/notes\\
\hline
{\bf Input values: \hskip 40pt}&&\\
Nominal H magnitude     & H=22 mag      &size: $\sim$ 140~m\\
NEO limiting magnitude  & 20.49 & \\
NEO distance    & 0.362 &au \\
Transverse velocity     &12     &km/s \\
Phase angle     &0      &deg \\
Telescope diameter      &100.0  &mm \\
Total QE        &0.64   & \\
Pixel size      &3.30   &\arcsec \\
Detector read noise     &1.20   &e$^-$\\
Frame time      &10.00  &sec\\
Total integration time  & 800.00        &sec\\
Total FOV       & 14.10 & $(\arcsec)^2$\\
Sky background  &22.0   &mag/(\arcsec)$^2$ \\
0V-mag reference        & $2.48\cdot 10^{10}$   &phot/m$^2$/s\\
\hline
{\bf Derived values: \hskip 40pt}&&\\    
Apparent magnitude      & 20.49 &mag\\
Flux detected   & 0.72  &e$^-$/s\\
Noise/frame variance & 83.86    & e$^-$\\
Signal/frame    & 7.17  &e$^-$\\
\hline
Total SNR       &7.00   &in 800~s\\
\hline
\end{tabular}
\end{scriptsize}
\end{center}
}
\vskip -20pt 
\end{table}
 
Table~\ref{tab:param} shows the sensitivity of a 10~cm synthetic tracking camera at 1~au. In calculating these sensitivities, we used the QE($\lambda$) of a commercial sCMOS detector\footnote{For details of the detector, please see: http://www.hamamatsu.com/jp/en/product/category/5000/5005/C11440-22CU/index.html} that peaks at 82\% at 0.55~$\mu$m. The QE was multiplied by a 5,800~K black body emission where 0.0 mag represented a flux of $9.66\times 10^9$~ phot/m$^2$/s into a 0.089~$\mu$m bandpass. This resulted in a detected flux of $2.48\times 10^{10}$~phot/m$^2$/s when integrated from $\sim$0.45--0.9~$\mu$m for a 0.0 mag NEO, which was assumed to have a solar like spectrum. While it would be reasonable to assume an optics efficiency of $\sim$80\%, to be conservative we assumed 55\%. We assumed the detector read noise to be 1.2e$^-$ and the zodi background at 1~au to be 22~mag/(\arcsec)$^2$.  The detector dark current at 0.006 e$^-$/pix/s at $-30~{}^\circ$C was ignored. The zodi background calculated was based on a solid angle that was calculated using the following information: the pixel size, the optical PSF of a Canon 400~mm f/4 lens with a $\sim$3.6\arcsec~ spot size in reflected sunlight, as assumed jitter of 1\arcsec~ in the spacecraft and a factor that is due to the PSF that can straddle more than one pixel. A commercial camera lens is designed to be focusable from 4~m to infinity. A custom lens designed only for focus at infinity could possibly be slightly better. The long integration time means that a NEO will move across several pixels and more or less uniformly sample the pixel. The effective background was calculated using a simulation of a moving object and a matched filter. The effective zodi background is from a 6.4\arcsec~ box, not quite $2\times2$~pixels when all the effects above are included (pixel size, pixel straddling, optical PSF and diffraction PSF).

If a camera moves closer toward the Sun, the relevant zodi background increases. In this regard, our simulation includes the noise contribution from the dust in the inner solar system by accounting for its intensity variation as a function of heliocentric distance \citep{Roach-Gordon:1973,Leinert-etal:1982}.

Compared to ground based telescopes, the numbers in Table~\ref{tab:param} may seem optimistic. However, many telescopes are not located at the darkest sites and even at the darkest sites, 21.5~mag/$(\arcsec)^2$ sky background is possible only at new moon. When the sky background due to a full moon gets to 18~mag/$(\arcsec)^2$, the loss in sensitivity ($\sim$2 mag) makes the search almost impossible.

\section{Emerging capabilities of interplanetary CubeSats}
\label{sec:cubesat} 

There is an on-going paradigm shift occurring in the satellite industry that may be compared to events dating back more than three decades to when personal computers disrupted mainframe computing. Small spacecraft, and their most popular sub-classification, CubeSats, have tremendous potential, not only in the commercial realm, but also by innovating established space programs through the use of CubeSats for research and technology development and demonstration. 

The low-cost CubeSat components, shorter development cycle, and availability of frequent launch opportunities for smaller satellites make it quicker and less expensive to get the latest capabilities into space.  JPL is involved in these efforts with several CubeSat projects both in LEO and for deep space, with several already launched and many more being developed.\footnote{For JPL's effort in the CubeSat area, please visit http://cubesat.jpl.nasa.gov/. The NASA efforts in this area are summarized in {http://www.nasa.gov/directorates/spacetech/ small\_spacecraft}} 

\subsection{System design: science instrument}

\begin{figure}[ht]
%\epsscale{0.5}
%\plotone{fig3.eps}
\centering\includegraphics[width=6.5cm]{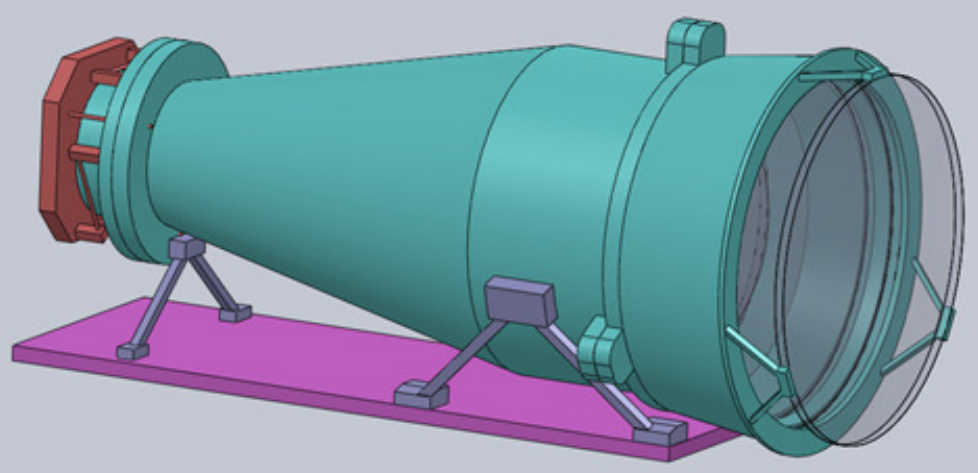}
\caption{A CAD design for the synthetic tracking camera for a CubeSat.
\label{fig:3}}
\end{figure}

The synthetic tracking camera is the primary instrument for the NEO search (Fig.~\ref{fig:3}).  The lens would be similar to commercially available camera lenses. The optics and structure would be designed to survive launch loads and will have the necessary thermal insulation to passively cool the instrument to maintain the optics temperature stable to within $5^\circ$~C of the desired value. 

The sensor is a 4K$\times$4K class sCMOS\footnote{For details on Scientific CMOS (sCMOS) cameras, please visit {http://www.andor.com/scientific-cameras/ neo-and-zyla-scmos-cameras}. Larger pixel format sCMOS detectors with 16~Mpix currently under development and are expected to be available in 2017.}  detector. Second generation sCMOS detectors have low $\sim1.2{\rm e}^-$ read noise and low $<0.1{\rm e}^-$/pix/sec dark current. The plate scale of the camera $\sim3.3$ \arcsec/pix, which sets the amount of zodi background per pixel.

\subsection{Observation cadence and data processing}
\label{sec:3.2}

A crucial capability for using synthetic tracking is to handle on-board data processing in real-time, which we now discuss. Our nominal observation uses exposure times of 10~sec and 800 sec integration time per field (80 frames). The detection threshold of SNR is set to 7. The slew time for moving from one field to another field is 10 seconds. With a FOV of $14({}^\circ)^2$, the sky can be scanned approximately for $\sim$13 days if we adopt the same approach as NEOCam to restrict the ecliptic latitude range to $\pm45^\circ$ and with sun exclusion angle beginning at $\sim60^\circ$. To determine the orbit of the detected asteroids, the observations of the same asteroid need to be connected. In Sec.~\ref{sec:discussion}, we will discuss how to arrange the observation cadence to make the connection of observations feasible. For most of our discussions and the case studies presented in Sec.~\ref{sec:case-studies}, we assume this simple sky scanning scenario.

We process our basic data set, which contains 80 images of size $\tt (4K \times 4K)$, in real-time (within 810~sec) using field-programmable gate array (FPGA). The main computation load comes from the synthetic tracking search, where we integrate 80 frames for each velocity vector in a two-dimensional $\tt (RA,DEC)$ velocity grid with spacing $\sim2$ pixels per 800~sec ($\sim$0.2~${}^\circ$/day). Because NEAs are typically detected at $\sim$0.36~au with a typical velocity of 10~km/sec, moving at $\sim$1~${}^\circ$/day, it is sufficient to search over a range of $\pm6~{}^\circ$/day, giving a velocity grid of size $(60 \times 60)$. The total amount of computations needed is estimated as $(4000)^2 \times 80 \times (60)^2$ per 810 sec or 5.7~GFLOPS. The signals detected by FPGA is then further analyzed by a four-parameter (2D location and velocity) least-squares fitting, which provides a better estimate of the SNR of the detected signal and more precise astrometry.

It is relevant to discuss the possibility of processing 800~sec of data that consists of two sets of 40 frames separated by 2~hrs without adding significant extra computation load. This is particularly useful for observation cadences supporting linkage of observations for orbit determination to be discussed in Sec.~\ref{sec:discussion}. A direct synthetic tracking search would require a velocity grid with a very fine resolution of 2 pixels per 2~hrs ($\sim$0.022 ~${}^\circ$/day) because the total temporal range of the data set is 2~hrs. To cover a velocity range of $\pm6~{}^\circ$/day, we need a velocity grid of size $540 \times 540$ giving $(4000)^2 \times 80 \times (540)^2$ operations per data cube (800~sec) or 467~GFLOPS to perform the synthetic tracking search.

We introduce a two-stage synthetic tracking to greatly reduce the amount of computations needed: 1) We perform synthetic tracking using a velocity grid with spacing of 2 pixels per 800~sec to separately integrate the first and second 40-frames to form two synthetic images; and 2) we combine the two synthetic images using a fine velocity grid with spacing of 2 pixels per 2 hrs.  For integrating 400~sec data, it is sufficient to use a velocity grid with spacing of 2 pixels per 800~sec ($\sim0.2~{}^\circ$/day). A grid of size $(60 \times 60)$ covers the range of $\pm 6~{}^\circ$/day. We estimate the amount of computation needed for the first stage to be $(4000)^2 \times 40 \times 2 \times (30)^2$ per data cube (820~sec, note there is one slew per 400~sec observation), or 5.6~GFLOPS. For the second stage, the amount of computation $(4000)^2 \times 2 \times (540)^2$ operations per data cube or $\sim$12~GFLOPS. We can further reduce the amount of the computation needed for the second stage by performing synthetic tracking only for pixels of interest.

Note that for a signal of SNR~$ = 7$ from integrating 80 frames, over the first 40 frames, we expect the SNR to be $\sim5$. We, therefore, can require that only pixels with SNR~$ > 3$ (using a lower threshold than 5 mitigates miss detection due to statistical fluctuation of the signal strength) from integrating the first 40 frames to be processed. This greatly reduces the amount of pixels to be considered for synthetic tracking by a factor of $\sim$700 (assuming Gaussian statistics). Thus the amount of computation needed for the second stage becomes $\sim$16~MFLOPS. So the total computation needed for synthetic tracking search is still $\sim$5.7~GFLOPS.

\subsection{Spacecraft design}
\subsubsection{LEO flight system from JPL's Team Xc}

A recent JPL's TeamXc study, a rapid concurrent design engineering session, developed a detailed design for a Low Earth Orbit (LEO) CubeSat with synthetic tracking \citep{Zarifian-etal:2014}.  The LEO flight system design relied on a 6U CubeSat architecture comprised largely of commercial components with flight heritage from previous CubeSat missions that have flown in LEO.  Here we describe all major components of the system, except for the telecommunication, navigation, and propulsion systems; these systems were augmented for the interplanetary application due to critical differences for an interplanetary destination and are discussed in the following Section. 
 
The avionics of the spacecraft will include the radiation-tolerant LEON processor\footnote{For details on LEON processors: {http://www.gaisler.com/}} and algorithms to perform data analysis, command, and control.  The computational requirements are dominated by multi-vector shift/add processing, which is required for the synthetic tracking algorithms. A top-level conceptual design of the computation architecture was studied using flight-qualified FPGAs. As discussed in Sec.~\ref{sec:3.2}, the processor requirement is $\sim$5.7~GFLOPS, which could be done by programming 24 computational units into a single flight-qualified FPGA. In addition to the FPGA, approximately 6~GB of RAM would also be needed to store the data for subsequent on-board processing. Approximately 10\% of the arithmetic processing capabilities of a Virtex-7 FPGA would be needed. The FPGA power consumption is anticipated to be $\sim$6--7~W.

The attitude control system will be based on the Blue Canyon Technologies XACT\footnote{For details on the Blue Canyon Technologies XACT unit: {http://bluecanyontech.com/product/xact/}. Also, for  XACT, High Performance Attitude Control for CubeSats, see {http://bluecanyontech.com/wp-content/uploads/2012/07/BCT-XACT-datasheet-1.5.pdf}} unit, which consists of a star tracker, reaction wheels, and an inertial measurement unit (IMU), to achieve the pointing stability and agility requirements of the mission. Based on the specifications, the XACT unit achieves pointing control of 10.8{\arcsec} (1-$\sigma$), knowledge of 6{\arcsec} (1-$\sigma$) and stability of 10.8{\arcsec} over 4 sec. The NEO CubeSat requires a pointing control at 2${}^\circ$ which is easily met by XCAT. However, the pointing stability requirement for NEO CubeSat is 6{\arcsec} over 4 sec, which is at the limit for the current XACT unit. Thus, some modification may be required to meet the pointing stability requirements once improved thermal/structural designs are available and an end-to-end simulation can be done. Such a modification may lead to additional isolation for jitter mitigation, which can be achieved by adding a fourth reaction wheel or other stabilization methods. 

A CubeSat reaction wheel, such as model RW8 by Blue Canyon technologies, has a max torque of 0.007~Nm, weighs 250~g and can slew a 10~kg spacecraft (30~cm max dimension) by $4^\circ$ in $< 7$~sec. We use 10~sec as a ``typical'' time to slew the telescope between adjacent FOVs.

The power system consists of i) deployed solar arrays that generate $\sim$40~W at the beginning of the mission, ii) a solar array drive assembly to point the arrays at the Sun, iii) the on-board battery to support high-powered events, and iv) a power management system. A standard Aluminum CubeSat structure will protect components from radiation in the LEO environment. A standard deployment system will house and deploy the spacecraft.  

The total current best estimate for a ``dry'' mass of the LEO system was estimated to be 8.2~kg, without margin \citep{Zarifian-etal:2014}. The cost, including mission and science operations, was found to be $\sim$\$9M, including 20\% margin.

\subsubsection{Computing architecture}

In developing the required computing architecture, we account for the fact that an average H=22 mag NEO can be detected by a 10~cm synthetic camera at a distance of $\sim$0.36~AU, when the observatory is at $\sim$0.9~au from the Sun.  If the NEO moves with a velocity of 12~km/s relative to the observatory, its angular motion is $\sim$0.046 {\arcsec}/sec. Since the pixel is 3.3{\arcsec}, we have to record images faster than one per 100 sec for the streak related loss in sensitivity to be small, for the total 800 sec of observation. We have chosen 10 sec exposures to capture i) NEOs with higher velocity and ii) smaller NEOs that can only be detected at ranges much closer than $\sim$0.36~au. A 800~sec observation would have 80 images.  When we shift/add these images, we should shift and add with a velocity vector range of velocities that includes the highest velocities a NEO can have and the spacing between velocity vectors should result in a velocity mismatch of less than 2 pixel/800 sec data cube. 

At a high level, the on-board data processing computer performs four core functions: data reduction and star removal, integer shift-and-add, candidate selection, and postage-stamp image generation for down-link. The shift-and-add operation provides all driving requirements, as it applies two orders of magnitude more arithmetic operations than data reduction and star removal, and the final two steps operate on significantly smaller datasets. 

\begin{figure}[ht]
%\epsscale{0.8}
%\plotone{fig4.eps}
\centering\includegraphics[width=9.0cm]{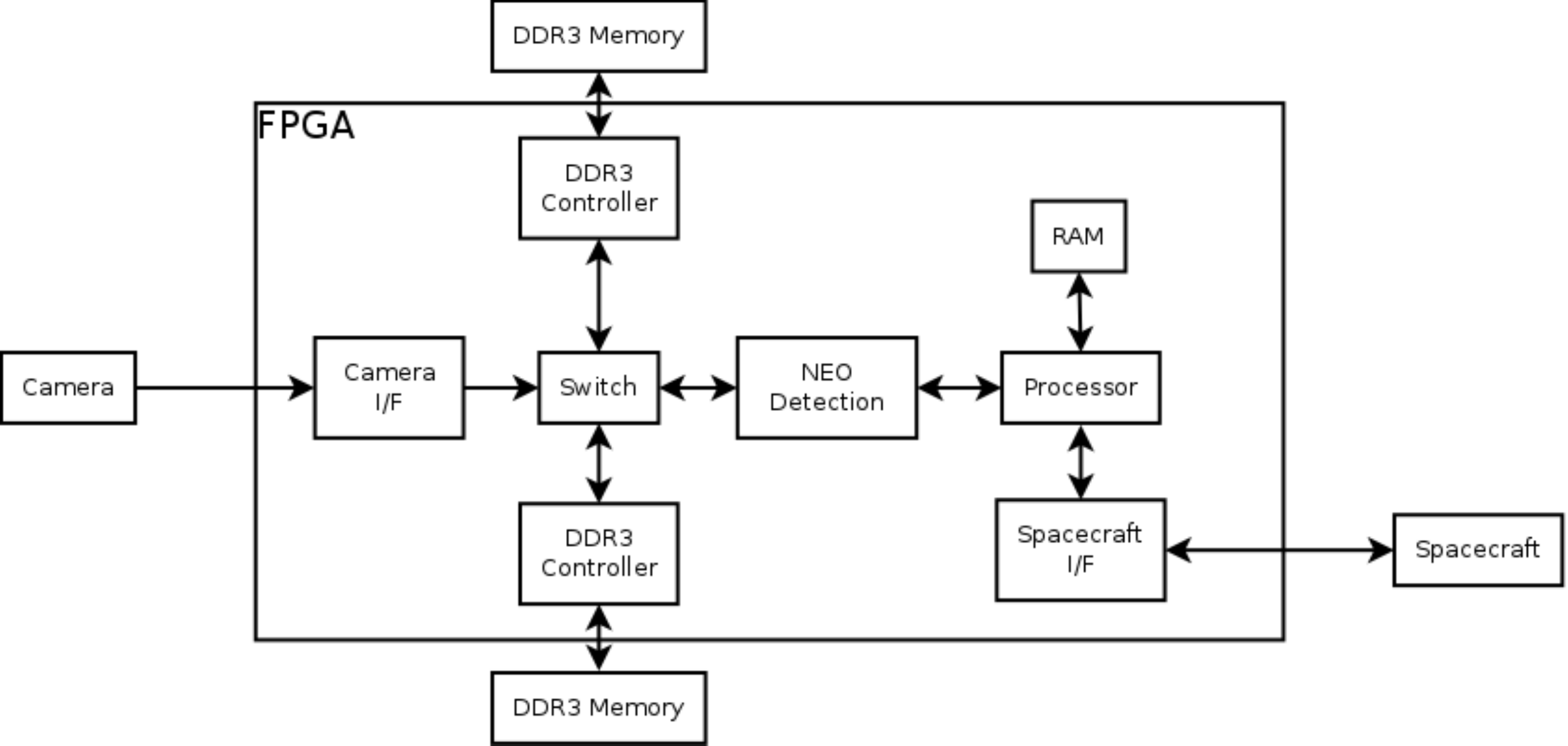}
\caption{Proposed FPGA-based computational architecture.
\label{fig:4}}
\end{figure}

Figure~\ref{fig:4} outlines our proposed FPGA-based computational architecture. We have baselined a Xilinx Virtex-7 device\footnote{For more information on Xilinx Virtex-7 devices, please see: {http://www.xilinx.com/products/silicon-devices/ fpga/virtex-7.html}}, which is currently slated to fly on the NASA TESS mission\footnote{For information on the NASA's Transiting Exoplanet Survey Satellite (TESS) mission, please visit {http://tess.gsfc.nasa.gov/}}. Camera data stream into the FPGA and are offloaded into one of the two separate external memory banks, which operate as a ping-pong buffer pair. Once an entire datacube is collected, data collection immediately resumes and targets the second memory bank. Data processing is performed as described in the previous section, with candidate objects reported to an on-board softcore microprocessor. The processor runs software to identify unique objects among the candidates and coordinates postage stamp collection and downlink to Earth.

For this initial design, we have chosen 3~GB of 64-bit DDR3 at 200~MHz per external memory bank, which supports a data rate of 12.8 GBps. By utilizing a parallel NEO detection algorithm with thirty separate streams running at 200 MHz and 1 operation per cycle, the system provides 6~GFLOPS.
 
A prototype implementation including memory controllers, shift-and-add logic, softcore microprocessor, and on-board memory has been built and run through FPGA design/simulation tools. The multivector-shift/add part of the processing uses $\sim$5\% of the arithmetic units available on  Virtex-7. This design was also analyzed with Xilinx's power estimation tool, with a resultant power draw of approximately 6.5 W split equally among configuration logic, device I/Os, and design logic.

\subsubsection{Required changes from the LEO to interplanetary NEO SmallSat design}

The interplanetary application of the NEO SmallSat requires changes to the LEO CubeSat design discussed above. The required changes to the hardware may be inherited from many common spacecraft elements developed for interplanetary CubeSat missions that have been or are currently being developed at JPL. These missions include INSPIRE\footnote{For information in INSPIRE Mission: {http://cubesat.jpl.nasa.gov/projects/inspire/overview.html}},  LunarFlashlight, NEAScout, and MarCO\footnote{For information on MarCO mission: {http://space.skyrocket.de/doc$\_$sdat/marco.htm}}, which are expected to fly before this NEO SmallSat, providing additional maturity and flight heritage to our design.  The major changes for the interplanetary CubeSat design will affect several subsystems including navigation, communication, and propulsion, as summarized in Table~\ref{tab:2}.

To enable navigation and telecommunication in deep space, the interplanetary NEO SmallSat will include the Iris transponder\footnote{http://mstl.atl.calpoly.edu/~bklofas/Presentations/DevelopersWork shop2014/ Duncan$\_$Iris$\_$Deep$\_$Space$\_$Transponder.pdf}  \citep{Duncan-etal:2014} and a High Gain Antenna (HGA), such as the one used on the NEAScout mission\footnote{For details on the NEA-Scout (Near Earth Asteroid Scout) is a CubeSat mission: {http://space.skyrocket.de/doc$\_$sdat/nea-scout.htm}}. With the Iris transponder and HGA with direct Earth transmission to the NASA's Deep Space Network (DSN), data rates of approximately 1~kbps will be achievable at the closest expected distance of 0.3~au (once the constellation is established), data rates of $\sim$90 bps are achievable at distances of 1~AU, and less than 30~bps at the maximum distance of 2~au.  Throughout the mission, the data rates will be adjusted according to the spacecraft distances from Earth.  To return the total of approximately 128 MBytes of data throughout the 4 year mission with six SmallSats, each spacecraft will download for 0.1\% or 16.4~mins/week (at 0.3 au) and 7\% or 12.2 hrs per week (at 2 au) of the mission time, which is accounted for in the engineering time allocation.   In future studies we will also investigate the option of sending data through inter-satellite relays to return the data to Earth to achieve higher throughput.

Synthetic tracking is sensitive to ``moving'' objects, especially to those that move over 3\arcsec~in 400 sec.  The {\tt neo.jpl.nasa.gov} site lists $\sim$100 ``near'' encounters of known asteroids. For the majority of these EGs (i.e., $\sim$98--99\%) their geocentric velocity is larger than 3~km/s. Synthetic tracking is immune to false positives from cosmic rays, galaxies, and stars near the detection threshold of 7-$\sigma$ (see Sec.~\ref{sec:syn-track_snr} for discussion). Although the Gaia mission can detect $\gtrsim60-80$\% of NEOs, it is capable of detecting all the main belt asteroids brighter than V=20 (or $\sim$2~km in size, see \cite{Tanga-Mignard:2012}).  Should we detect a moving object from the Gaia catalog, we will not downlink the corresponding 125 pixel postage stamp of that object nor will we follow it up with observations 2 hrs later.  This minimizes the down-link data to $\sim$20,000 NEOs brighter than $\rm H=22$~mag. Our simulations show that, on average, each of these NEOs will be detected approximately ten times during the 3.5 year mission. The ultimate telecom system for our mission would allow for this anticipated data volume. Fitting orbits with only ten detections is challenging, but some progress in this area has been made  \citep{Granvik-Muinonen:2005,Granvik-etal:2007,Granvik-Muinonen:2008}; we will use these techniques in the future.

\begin{table}[h!]
\begin{center}
\begin{scriptsize}
\caption{The cost estimate for the interplanetary version of NEO SmallSat resulting from changes to a LEO CubeSat design.\label{tab:2}}
\vskip -5pt
\begin{tabular}{|p{3.28cm}|p{3.17cm}|c|}
\hline\hline
Change & Function & Cost, \$M\\
\hline
\multicolumn{2}{|r|}{Total LEO CubeSat (before margins)} &7.5\\\hline
Replacement of UHF radio with Iris deep-space Transponder &
To enable interplanetary communication and navigation   &  0.4\\\hline
Addition of propulsion system   & Busek Thrusters (4x)  &  1.0\\\hline
Additions to ACS, power, structure, thermal subsystems &        Increased spacecraft size, mass, and design work expected &  0.5\\\hline
Operations      &Trajectory planning and ground systems   &1.0\\\hline
\multicolumn{2}{r|}{Total change} &2.9\\\cline{3-3} 
\multicolumn{2}{r|}{Total Interplanetary NEO SmallSat} 
&10.4\\ 
\multicolumn{2}{r|}{(before margins)} 
&\\\cline{3-3} 
\multicolumn{2}{r|}{Total Interplanetary NEO SmallSat}& 12.5\\
\multicolumn{2}{r|}{ (including 20\% margin)}& \\
\cline{3-3} 
\end{tabular}
\end{scriptsize}
\end{center}
\vskip -10pt
\end{table}

To send the SmallSat into a solar orbit at heliocentric ranges of less than 1 AU, a dedicated propulsion system is required. Compared to the ground-based observations or a LEO version, the solar orbit would provide the mission with a better observing geometry and a faster orbit to conduct NEO/EG search. The CubeSat would share a ride to GEO and then it would use an extra propulsion motor to enter solar orbit. Although there are no  currently existing CubeSat-class propulsion modules with the required capabilities, several promising propulsion options are available, including the Busek Electrospray Thrusters (BET)\footnote{For details on the Busek Electrospray Thruster system, please visit http://busek.com/technologies$\_\_$espray.htm} and chemical thrusters built by Aerojet Rocketdyne\footnote{For details on the Aerojet Rocketdyne thruster, http://www.rocket.com/cubesat}. There are other propulsion options based on the systems that are currently being developed. In the near future these new systems will be able to satisfy the size, power, thermal, and launch constraints of a small spacecraft and should be considered in an end-to-end system optimization \citep{Mueller-etal:2010,Marrese-Reading-etal:2010,Spangelo_JSR2014}.

We also expect changes to the power, thermal, structural, and attitude control systems. These changes will result in increases in mass, volume, power, and cost estimates for the EG SmallSat, as reflected in Table~\ref{tab:2}.  The highest power for the interplanetary CubeSat is expected to be approximately 47 W during the download mode. To accommodate these changes we would have to resize the solar panels and/or batteries.  Trajectory planning and ground operations are also expected to result in additional cost increases.  Although pointing requirements are similar to the LEO version, we expect to use larger reaction wheel assemblies (RWA) which may result in larger torques on the spacecraft during RWA de-saturation and motivate changes to the algorithms and thruster controls during RWA de-saturation periods.  Additional spacecraft structural thickness will be implemented surrounding the radiation-sensitive components such as components of the attitude control system (ACS)  (i.e., the stellar reference unit (SRU), RWA driver) and the electrical power subsystem (EPS). The small
spacecraft community is developing other safety and mission assurance strategies  to mitigate these concerns for interplanetary missions \citep{Sheldon-etal:2014}.  

The resulting spacecraft will be a 9U CubeSat bus. The cost estimate for the interplanetary NEO CubeSat is \$12.5M, including a 20\% margin, as shown in Table~\ref{tab:2}. We start with a cost estimate for the LEO CubeSat provided by the JPL Team Xc and add our estimates for the cost changes for each of the affected subsystems. Table~\ref{tab:2} shows the cost estimate for a spacecraft that could be used for a NEO SmallSat constellation. It shows that a constellation capable of detecting 90\% of EGs would be 5--15 times less expensive than the missions/facilities considered in the 2010 NRC report. A more detailed mission design study would be needed to refine the cost and will be done in the near future.

\subsection{Spacecraft mass and power budgets}
\label{app:b4}

The CubeSat is expected to have a dry mass of approximately 13.2 kg, including payload and all major subsystems power, propulsion, command and data handling, attitude determination and control, thermal, and structure. Table~\ref{tab:3} shows the current best estimate (CBE) for the masses of each subsystem.  The required propellant mass will depend on the orbit chosen for the spacecraft, and is expected to be $\sim$9 kg (to boost to an orbit of $\sim0.8$ au). Therefore, the total expected wet mass is approximately 22.2~kg before margins, representing realistic mass allocation for a SmallSat to be flown on an interplanetary mission.

\begin{table}[h!]
\begin{center}
\begin{scriptsize}
\caption{The current best estimates (CBE) for the mass of the interplanetary NEO SmallSat.\label{tab:3}}
\vskip -5pt
\begin{tabular}{|p{2.8cm}|p{3.4cm}|c|}
\hline\hline
Subsystem & Components & Mass, kg\\
\hline
Science instrument      & Synthetic tracking camera &     2.8 \\\hline
Navigation, communication       & Iris transponder;
high gain antenna & 1.0 \\\hline
Command, data handling 
& LEON processor and board        & 0.25 \\\hline
Attitude control system 
        & BCT XACT with sized-up wheels  & 0.85 \\\hline
Propulsion &    Busek thrustes (4x)       &4.8 \\\hline
Thermal & Radiator        &1.0 \\\hline
Power   & Solar panels, batteries, EPS    & 0.86 \\\hline
Mechanical      & Structure       &2.0 \\\hline
\multicolumn{2}{r|}{Total Dry} &13.2\\\cline{3-3} 
\multicolumn{2}{r|}{Total Wet (without margin)}& 22.2\\
\cline{3-3} 
\end{tabular}
\end{scriptsize}
\end{center}
  \vskip -25pt
\end{table}

\subsection{Spacecraft lifetimes and constellation architectures}
 
\begin{figure}[t!]
  \begin{center}
%\epsscale{0.45}
%\plotone{fig5.eps}
\centering\includegraphics[width=6.3cm]{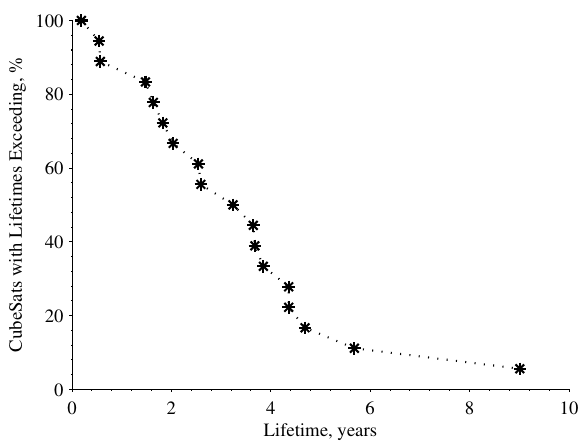}
\caption{Lifetime Statistics from historic LEO CubeSat missions (filtered for unlikely failures).
\label{fig:5}}
  \end{center}
  \vskip -20pt
\end{figure}

The ability to achieve the mission's science objectives of detecting 90\% of all the EGs is directly related to the mission lifetime; therefore, we are concerned about CubeSat failures. We have studied historical data on CubeSat failures, available for missions developed and launched by universities, government, and industry. Figure~\ref{fig:5} shows the statistics for typical lifetimes of over 200 CubeSats (including 1U, 2U, 3U) launched since 2003, where the full data set is described by \cite{Sheldon-etal:2014}. We closely examined the data and filtered out failures due to launch or deployment, and those due to causes that are expected to be preventable such as communication or power problems due to poor designs, or latch-up due to potential radiation exposure.  

We also filtered out CubeSats that de-orbited, where they did not fail due to technical reasons, and those that were only recently launched and have not yet failed. Most of this historical data is based on university-built CubeSats, so using this data to inform expected lifetimes is conservative as the proposed CubeSats will be developed by professional engineers using high-heritage components. Furthermore, there is no available statistical information about planetary CubeSats because they have not yet been launched, so this is the only data that we can use to extrapolate performance. Assuming a Gaussian distribution, the lifetime is $\approx 2.1$~yrs. Based on our statistics, $\approx$45\% of CubeSats have lifetimes over 3.5 years, as shown in Fig.~\ref{fig:5}. This indicates the trend in technological maturity and growth of the lifetimes of the CubeSats. Two MarCO CubeSats\footnote{For more information on MarCO CubeSats, pelase see {http://www.jpl.nasa.gov/cubesat/missions/marco.php}} have been completed, and are waiting for launch. They are designed to piggy back on a ride to Mars and act as communication relay satellites.
  
\subsection{Setting up a constellation of SmallSats}
\label{sec:constellation}

A detailed mission analysis that optimizes the mass and time of setting up a NEO SmallSat constellation, and associated cost is beyond the scope of this paper. Here we take a simplified approach to describe a sequential transfer of several SmallSats from an Earth's orbit to a 0.8 au orbit. We start with the first spacecraft. A spacecraft in a 0.8 AU orbit has a period of 261 days. After $\sim2.5$ years, it will have traveled around the Sun $\sim1,260^\circ$, while the Earth will have traveled $\sim900^\circ$ around the Sun. A low cost way to deploy a constellation of six SmallSats to a heliocentric orbit would be to send them out from the Earth's orbit every $\sim152$ days. The time between the 1st of the six spacecraft and the last would be $\sim2.1$ years.

For constellation orbits close to 1 au, for example, 0.95 au, this method of ``setting up'' a constellation can be time consuming. With an additional propulsion, one could enter an orbit with a smaller semi-major axis of say,  0.8 au, wait for the spacecraft to race ahead of Earth, then fire the thruster to get back to 0.95 au. Most ($\sim70$\%) of the 9 kg of propellant mentioned in Appendix~\ref{app:b4} is used to escape Earth's gravity (from LEO), so more complex orbit maneuvers would not be prohibitively expensive.

Since the SmallSats can observe as they transit to the 0.8 au orbit, the transfer time could, in principle, lengthen the mission lifetime by  only $\sim1.1$~years.  The simulation of six 9U CubeSat in Sec.~\ref{sec:case-studies} showed that $\sim$3.8 yrs of observations could detect 90\% of H = 22~mag NEOs. Therefore, at least in theory, the survey could be conducted in $\sim$5~years after the 1st spacecraft leaves the Earth's orbit.  Our preliminary analysis has shown that the Congressionally mandated NEO survey goals may be accomplished in $\sim 5$ years by a constellation costing a small fraction of the missions studied in \citep{NRC2010}. This is the unique advantage of a SmallSat-based mission architecture, which enables trade-offs to be made regarding the survey completion time, risk, and cost.  

The \citep{NRC2010} report concluded that it would take approximately ten years for a single spacecraft with a 50~cm telescope to conduct a search for 90\% of H=22 mag NEOs. Designing a spacecraft with the redundancy and required testing to ensure survival for ten years can significantly increase the mission cost. With a constellation of six SmallSats, a failure of a single spacecraft is not catastrophic; in fact, the constellation's performance degrades gracefully with respect to loosing a node.  Fig.~\ref{fig:11} shows the sensitivity in search time to constellation size.  Given the low cost of an additional SmallSat, one may consider placing more than six spacecraft at the desired solar orbit to form a redundant constellation.  Such a redundancy would not only reduce the time of conducting the 90\% NEO survey, it would also reduce the risk of the NEO search to potential spacecraft failures identified in \citep{NRC2010}.  

\end{appendix}

\end{document}